%% file: main.tex
\documentclass[12pt]{article}
\usepackage[utf8]{inputenc}
\usepackage{epsfig}

\usepackage[letterpaper]{geometry}
\usepackage{booktabs}

\usepackage{graphicx}
\usepackage[usenames,dvipsnames]{color}
\usepackage{todonotes}
\usepackage{bbm} 

\usepackage{tikz}
\usetikzlibrary{topaths,calc}

\newcommand{\citep}[1]{{\cite{#1}}}

\newcommand{\eps}{\varepsilon}

\newcommand{\R}{{\mathbb R}}

\usepackage{graphicx}
\usepackage[colorlinks=true, allcolors=blue]{hyperref}

\usepackage{paralist}
\usepackage{amsmath}
\usepackage{amssymb}
\usepackage{amsthm}
\usepackage{algorithmic}
\usepackage{algorithm}
\usepackage{tikz}
\usetikzlibrary{arrows}
\usetikzlibrary{arrows.meta}

\usepackage[]{todonotes}   
\usepackage{soul}

\newcounter{cc}
\newcommand{\ccc}{\thecc \addtocounter{cc}{+1}}

\author{
Stan Matwin\thanks{Faculty of Computer Science, Dalhousie University, Halifax, NS, Canada and Institute of Computer Science, Polish Academy of Sciences, Warsaw, Poland; e-mail: \texttt{stan@cs.dal.ca}}
\and
Aristides Milios\thanks{Faculty of Computer Science, Dalhousie University, Halifax, NS, Canada; e-mail: \texttt{amilios@dal.ca}}
\and
Pawe\l{}~Pra\l{}at\thanks{Department of Mathematics, Ryerson University, Toronto, ON, Canada; e-mail: \texttt{pralat@ryerson.ca}}
\and 
Amilcar Soares\thanks{Department of Computer Science, Memorial University of Newfoundland, St. John's, NL, Canada; e-mail: \texttt{amilcarsj@mun.ca}}
\and
Fran\c{c}ois Th\'eberge\thanks{Tutte Institute for Mathematics and Computing, Ottawa, ON, Canada; email: \texttt{theberge@ieee.org}}
}

\title{Survey of Generative Methods \\ for \\ Social Media Analysis\thanks{We acknowledge the support of the Communications Security Establishment and Defence Research and Development Canada. The scientific or technical validity of this report is entirely the responsibility of the authors and the contents do not necessarily have the approval or endorsement of the Government of Canada.} 
}

\begin{document}

\maketitle

\newpage

\tableofcontents

\newpage

\include{all_text}






\bibliographystyle{abbrv}
\bibliography{refs}




\end{document}

%% file: all_text.tex
\section{Introduction} 

This survey draws a broad-stroke, panoramic picture of the State of the Art (SoTA) of the research in generative methods for the analysis of social media data. It fills a void, as the existing survey articles are either much narrower in their scope~\cite{alexandridis2021survey} 
or are dated~\cite{batrinca2015social,stahl2014survey, yue2019survey}. 
We included two important aspects that currently gain importance in mining and modelling social media: dynamics and networks. Social dynamics are important for understanding the spreading of influence or diseases, formation of friendships, the productivity of teams, etc. Networks, on the other hand, may capture various complex relationships providing an additional insight and identifying important patterns that would otherwise go unnoticed.

The article is  divided in five chapters and provides an extensive bibliography consisting of more than 250 papers. Open problems, highlighting potential future directions, are clearly identified. We chose sentiment analysis as an application providing common thread between the four parts of the survey.

We start with Chapter \ref{chp:ontologies} devoted to the discussion of data models and ontologies for social network analysis. We organized the data models based on the concepts they use to solve a social media research problem such as homophily, social identity linkage, and personality analysis. 
We also discuss some ontologies for sentiment analysis and situational awareness. We conclude this chapter with highlighting promising research directions such as working with metadata and federated learning. 

Chapter \ref{chp:textgen} is devoted to text generation and generative text models and the dangers they pose to social media and society at large. The current SoTA in text generation, i.e.\  large pre-trained autoregressive Transformer models, the prime example of which is GPT3, is highlighted. These models are trained on massive amounts of data (e.g.\ Common Crawl), and have hundreds of billions of parameters. This allows them to generate eerily coherent text that is near-indistinguishable from text written by humans. The potential of these models for nefarious use is outlined, and potential ways to mitigate these harms via ``fake news'' detection through contextual information are provided as well.

Chapter \ref{chp:topsent} is devoted to topic modelling and sentiment analysis in context of social networks. Traditional topic modelling approaches are briefly described. Following this, methods to fuse deep learning and these traditional approaches for topic modelling are outlined in detail. The unique challenges that social media content poses for both topic modelling and sentiment analysis, as well as approaches that seek to mitigate them are discussed. In terms of sentiment analysis, both unsupervised rule-based approaches and transfer-learning-based approaches using large Transformer models (the current SoTA for complex sentiment analysis) are discussed. Finally, some interesting developing fields within sentiment analysis are outlined, with regards to both multimodal and target-based sentiment analysis.

Chapter \ref{sec:networks} is devoted to graph theory tools and approaches to mine and model social networks. Such tools become increasingly important in machine learning and data science. There are many important aspects so we tried to narrow the topics down to a few most important ones. We concentrate on graph embeddings and their evaluation, higher order structures, dynamics, and synthetic models.

\newpage

\section{Ontologies and Data Models for Cross-platform Social Media Data}
\label{chp:ontologies}

The creation of social media platforms generated an immense volume and diversity of content produced and exchanged by users. 
According to \cite{johnson2020semantically}, Facebook, Twitter, and Instagram boast over two billion monthly active users, and as such, their ability to directly and indirectly connect the world's population has never been easier or more far reaching.
Also, according to a Pew Research study \cite{pewRS2016}, 56\% of US adults online use more than one social media platform.
The plethora of heterogeneous online platforms fostered a vast scientific production on various applications of social media analysis, ranging from sentiment analysis to cyber influence campaigns. 
Integrating data from different social media platforms is challenging because many of them were created for multiple purposes. 
For example, while Linkedin is mainly focused on professional networking and career development, Twitter is used in diverse ways by different groups of users as stated in \cite{honey2009beyond}. 
The way users interact and produce content in such platforms are also heterogeneous and include likes, dislikes, shared videos or images, voting, friendships or connections, posts, and private messages. 
In this chapter, we discuss data models to organize the social media data by topics of common users' interests (Section \ref{sub:datamodels}) or the use of ontologies to organize data  from heterogeneous sources (Section \ref{sub:ontologies}). 
In all subsections, we detail and discuss the most relevant works (i.e., with a greater number of citations over the years) in the aforementioned topics.

\subsection{Data Models for Social Media Data Analysis}
\label{sub:datamodels}

In this section we discuss approaches for merging social media data with external sources.
For example, several approaches propose to enhance tweets or other social media data sources by annotating them with unambiguous semantic concepts defined in external knowledge bases such as Wikipedia or DBpedia. 
These knowledge bases provide an explicit semantic representation of concepts and their relations.
They, therefore, provide additional contextual information about tweets and their underlying semantics, allowing the creation of group of users with similar topics or interests.

The annotations used by most of the works that use twitter data are either provided by its  API\footnote{https://developer.twitter.com/en/docs/twitter-api/annotations/overview} or are done with online and crowdsourced tools\footnote{https://www.lighttag.io/}\footnote{https://www.tagtog.net/}\footnote{https://github.com/doccano/doccano}, or tools such as GATE \cite{bontcheva2013gate} or Webanno \cite{yimam2013webanno}. 
Most tools work by allowing users to select a word or a sentence and tag it with a given value. 
A good and recent reading on the annotation topic can be found in \cite{daudert2020web}.
The reliability on the data mainly depends on the expertise of the annotators used in the tagging process and in the number of tags provided to the same instance (i.e., tweet). 

We divided the data models based on the concepts they use to solve a social media research problem. 
We first discuss works that use the concept of homophily, and then move on to the social media linkage problem.
Finally, we show some works that uses images to infer personality traits of users. 

\subsubsection*{Homophily Analysis}

Homophily is the tendency of individuals to befriend other individuals sharing the same interest in Twitter communities \cite{faralli2015large}. 
Modeling the perception of friendship to perform homophily analysis may be challenging. 
A dataset enriched with the user's activity or interests is necessary to measure homophily since the social graph itself does not contain such information. Generally, the works in this area would use text from the messages (e.g., topic modeling) or some meta information of the social  (e.g., the similarity of the time-zone, popularity, user's subgraph in the vicinity, etc).
The notion of homophily has also commonly been modeled in social networks by mutual-follow, and mutual-mention relations \cite{barbieri2014follow}.

\begin{figure}[ht]
\centering
\includegraphics[width=.65\textwidth]{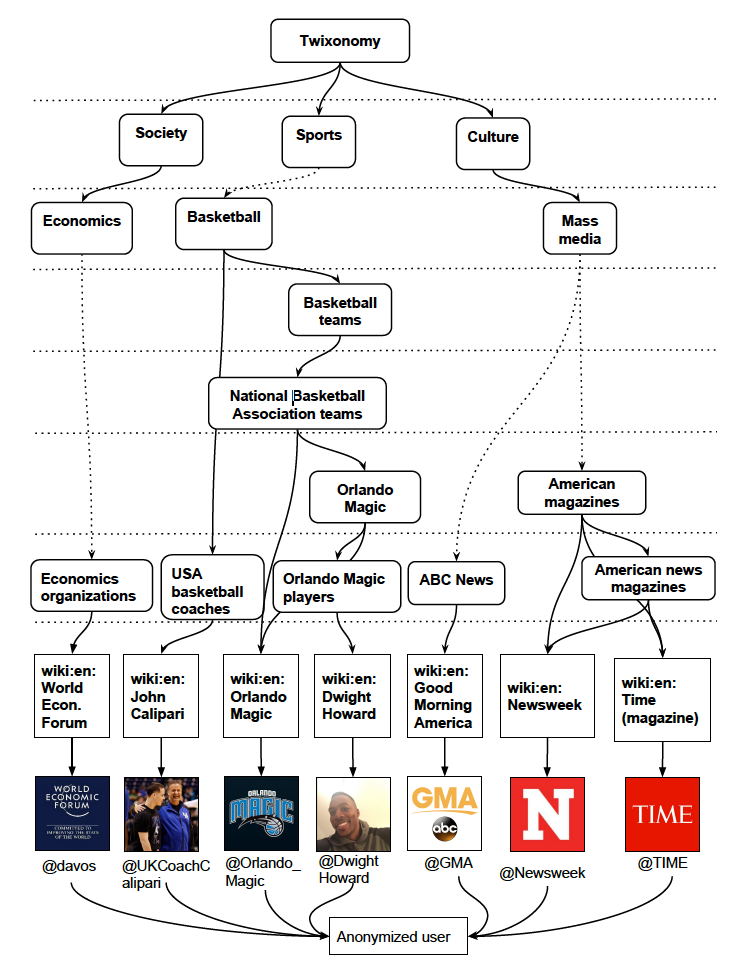}
\caption{Twixonomy example. Source:  \cite{faralli2015large}.}
\label{fig:twixonomy}
\end{figure}

Paper \cite{faralli2015large} proposes Twixonomy, which is a  novel method for analyzing homophily in large social networks based on a hierarchical representation of users’  interest.
The outcome of Twixonomy is a Directed Acyclic Graph (DAG) taxonomy where leaf nodes are pages from Wikipedia associated to Twitter  users per topic, and the remaining nodes are Wikipedia categories.
The authors associate Wikipedia pages with topical users in users' friendship lists to obtain a hierarchical representation of interests.  
Many pages can be associated with a user name, and to handle such a problem, they use a word sense disambiguation algorithm. 
Users can then be directly or indirectly linked to one or more Wikipedia pages representing their interests. 
The algorithm starts from a set of wikipages representing users' interests, and they consider Wikipedia categories as a sub-graph induced from these pages. 
Cycles are removed to obtain a DAG, and efficient cycle pruning is also performed using an iterative algorithm. 
The advantages of the Twixonomy include a compact, tunable and readable way to express the users' interest and it uses only interests explicitly expressed by the users. 
Figure \ref{fig:twixonomy} shows the Twixonomy of a ”common” user with 7 topical friends in his/her friendship list. 
Wikipages are the leaf nodes of the Twixonomy in Figure \ref{fig:twixonomy}, and the other nodes are Wikipedia categories layered by generality level. 
The mid-low categories are the most representative of a user’s interests since, as the distance between a Wikipage and a hypernym node increases, the semantic relatedness decreases \cite{faralli2015large}. 
In the example, the categories Economics, Basketball, and Mass Media could be chosen to summarize all the user’s primitive interests \cite{faralli2015large}. 
The experiments performed in \cite{faralli2015large} shows that while homophily is indeed a significant phenomenon in Twitter communities it is not pervasive. 
The authors conclude that inferring  user’s preferences on the basis of those of their friends is not a fully reliable strategy.
In a second experiment, the authors show that homophily also depends to some extent on the interests that identifies a community \cite{faralli2015large}.
The results show that people interested in education and fashion are more homophilous and, at the same time, those supporting political leaders and women’s organizations have a minor tendency to befriend other users with the same interests. 

Twixonomy is used in \cite{faralli2015women} for examining the distribution of interests in Twitter according to gender. 
Paper \cite{faralli2015women} uses a large list of female and male names extracted from several sources to classify the gender, and they also analyzed two populations: common users and topical users. 
The results showed that the proportion of celebrities and peers’ interests in the topmost categories is not statistically significant than the respective ratio in whole populations, except for the category Sports, where males dominate \cite{faralli2015women}.
The experiments also found very few women leaders, but women are indeed interested in leadership, but it seems that they prefer to follow male leaders.
Also, men have a significantly higher tendency towards homophily than women.
The experiments also point out that except for the categories Writers, Democrats, and Women’s organizations, women are either non-homophylous or support man or non-gendered entities significantly more than other women \cite{faralli2015women}.

\subsubsection*{Social Identity Linkage}

Social identity linkage is the problem of linking users across different social media platforms. 
A survey on this topic can be found in \cite{shu2017user,xing2019survey}.
The objective is to obtain from social media data a deeper understanding and more accurate profiling of users. 
There are several applications to be built from linking user identities, such as enhancing friend recommendations, information diffusion, and analyzing network dynamics.

\begin{figure}[ht]
\centering
\includegraphics[width=.7\textwidth]{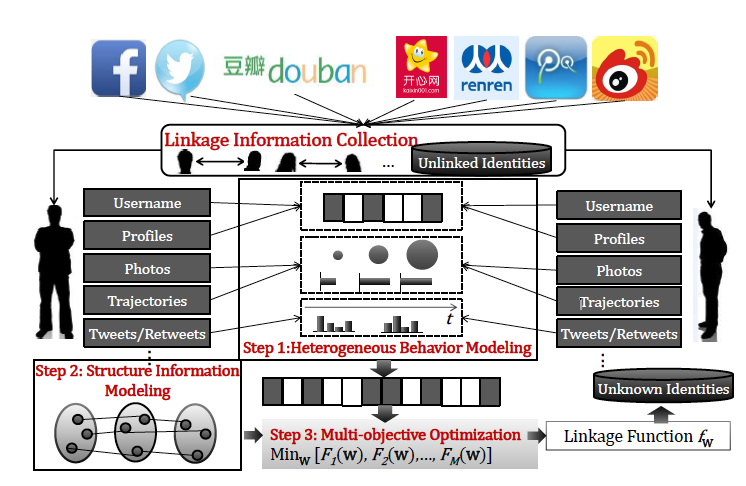}
\caption{Hydra framework. Source:  \cite{liu2014hydra}.}
\label{fig:hydra}
\end{figure}

Paper \cite{liu2014hydra} proposes a framework  
for cross-platform user identity linkage via heterogeneous behavior modeling named HYDRA. 
HYDRA is divided into three steps and can be seen in Figure \ref{fig:hydra}.
First, in the behavior similarity modeling, the relationship between two users of a pair for all user pairs via heterogeneous behavior modeling is calculated. 
In the second step the framework builds a structure consistency graph on user pairs by considering both the core network structure of the users and their behavior similarities. 
Finally, a multi-objective optimization is done based on the previous two steps, which jointly optimizes the prediction accuracy on the labeled user pairs and multiple structure consistency measurements across different platforms.
HYDRA uses common textual attributes present in user profiles such as name, gender, age, nationality, profession, education and email account and visual attributes such as face images used in the profile.
The authors evaluated HYDRA against the state-of-the-art solutions on two real data sets — five popular Chinese social networks and two popular English social networks.  
In summary, they evaluated a total of 10 million users and more than 10 terabytes of data and results demonstrated that HYDRA outperformed other baselines in identifying true user linkage across different platforms.

Paper \cite{zhou2018deeplink} proposes a deep reinforcement learning  comprehensive framework to address the heterogeneity called DeepLink to study the social identity linkage problem. 
DeepLink is an end-to-end network alignment approach and a semi-supervised user identity linkage learning algorithm that does not require a heavy feature engineering and can easily incorporate features created from the users' profiles.
DeepLink takes advantage of deep neural networks to learn latent semantics of both user activities and network structure in an end-to-end manner. 
It also leverages a semi-supervised graph regularization to predict the context (neighboring structures) of nodes in the network.
The experiments conducted demonstrated that the proposed framework outperforms various user identity linkage methods in linkage precision and ranking matching user identity.

\subsubsection*{Personality Analysis}

Generally, sentiment analysis variables takes values such as positive, negative, or neutral. 
These variables can also have a more extensive range of values, allowing for multiple assignments of sentiment in a single word. 
Additional meta-features based on the sentiment values can also be generated, such as subjectivity and polarity. 
Subjectivity is the ratio of positive and negative sentences to neutral sentences, while polarity is the ratio of positive to negative sentences. 
This is a very active research area and recent survey with past and recent works in this topic, mainly with Twitter data, can be found in \cite{antonakaki2021survey}.

Several works have been done to predict personality traits from twitter data. 
For example, in \cite{golbeck2011predicting} the authors propose a method by which a user’s personality can be accurately predicted through the publicly available information on their Twitter profile (e.g., number of followers, followed, mentions, hashtags, replies, density of the social network, etc).
Three tools were used for feature generation \cite{golbeck2011predicting} with the objective of analyzing the content of users’ tweets:
the Linguistic Inquiry and Word Count (LIWC) tool \cite{pennebaker2001linguistic} (e.g., 81 features in five categories),  MRC Psycholinguistic Database (e.g., over 150,000 words with linguistic and psycholinguistic features of each word) and the General Inquirer dataset\footnote{http://www.wjh.harvard.edu/ inquirer/} (provides a hand annotated dictionary that assigns words sentiment values on a -1 to +1 scale).
The work \cite{quercia2011our} in shows a study on the relationship between personality traits and five types of Twitter users: listeners (those who follow many users), popular (those who are followed by many), highly-read (those who are often listed in others’ reading lists), and two types of influential. 
The work also tries to the predict a user’s personality traits out of three variables that are publicly available on any Twitter profile: the number of profiles the user follows, number of followers, and number of times the user has been listed in others’ reading lists. 
The results presented in \cite{quercia2011our} show that all user types (listeners, popular, highly-read, and influential users) are emotionally stable (low in Neuroticism), and most of them are extrovert.
Results also show that user personality can be easily and effectively predicted from public data, and openness is the easiest trait to predict, while
extraversion is the most difficult.

The rest of this section will focus on image-based personality analysis. 
Recent research shows that personality traits can be inferred based on image-based content analysis and a survey can be found in \cite{bilal2019social}. 
Pictures include many features such as objects, colors, faces that can be automatically extracted using modern computer vision algorithms. 
These features can be used to examine the relationships between users' personalities and image posting across different social media platforms. 
For example, images can be used for detecting users' anxiety and depression as shown in \cite{guntuku2019twitter}.
The authors explore how depression and anxiety traits can be automatically inferred by looking at images that users post and set as profile pictures. 
They compare different visual feature sets extracted from posted images and profile pictures. 
The analysis of image features associated with mental illness essentially confirm previous findings of the indications regarding depression and anxiety. Facial expressions of depressed users show fewer signs of positive moods, such as less joy and smiling, and appear more neutral and less expressive. 
Interestingly, depressed individuals' profile pictures are marked by the fact that they are more likely to contain a single face (i.e., user's face) rather than show the user surrounded by friends.

\begin{figure}[!h]
\centering
\includegraphics[width=.8\textwidth]{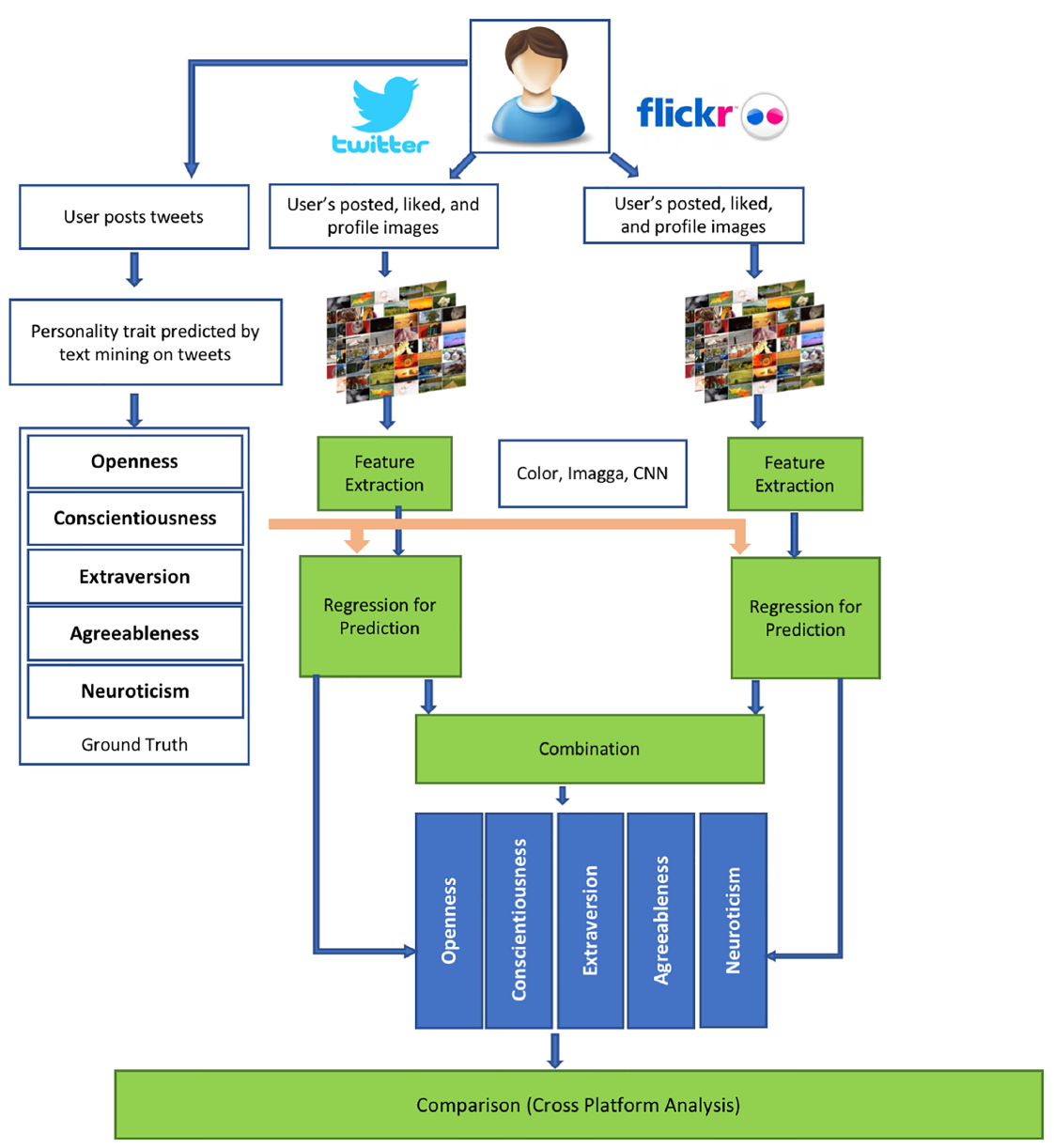}
\caption{Overview of cross-platform analysis for user personality prediction. Source:  \cite{samani2018cross}.}
\label{fig:personality-cross}
\end{figure}

The work shown in \cite{samani2018cross} tries to quantify image sharing preferences and to build models that automatically predict users' personality in a cross-modal and cross-platform setting using Twitter and Flickr. 
Figure \ref{fig:personality-cross} show the process of the cross-modal and cross-platform analysis. 
First, the authors assemble a dataset containing user posts, profile images, liked images and texts. 
After, they extract features from the images and analyze the text to predict personality traits (e.g., openness, conscientiousness, extraversion, agreeableness, and neuroticism).

The results presented in \cite{samani2018cross} show that multiple interactions that users have with social media platforms (i.e., choosing profile pictures, posting, and liking images) have predictive utility for automatic personality assessment of users. 
Predictive results are also boosted when information from multiple social networks are combined. 
Results show also that users' posted images had the best performance in predicting personality, followed by liked images and, finally, profile pictures. 
Liked images are more diverse in their content, and as result, algorithms would need a more extensive set of such pictures across the user's timeline to make more accurate predictions.

\subsection{Ontologies for Social Media Data}
\label{sub:ontologies}

A popular and reasonable choice to integrate heterogeneous sources such as social media data is by defining an ontology. 
An ontology represents the domain knowledge as a hierarchy of concepts \cite{gruber1995toward} and includes machine-interpretable definitions of the domain's basic terms, and relations \cite{guarino1998formal}. 
Ontologies also define a common vocabulary for researchers who needs to share information in a domain. 
Defining an ontology for a given problem or domain helps share a general understanding of knowledge among different teams and makes the domain knowledge reusable.
The next sections describe two main tasks related with social media data and ontologies which are sentiment analysis 
and situational awareness. 

\subsubsection*{Ontologies for Sentiment Analysis}


Sentiment analysis for subject information extraction from the text data has become more dependent on natural language processing methods, especially for business and healthcare, since online products and service reviews may affect consuming behaviors. 
A survey in multimodal sentiment analysis can be found here \cite{soleymani2017survey}.
Sentiment analysis algorithms typically apply natural language processing techniques with additional resources (e.g., sentiment and emotion based lexicons, sophisticated dictionaries, and ontologies) to model the documents. 
The Plutchik’s model \cite{plutchik1980general} is a common choice of several authors to assign labels that may reflect how users feel about topics, images, and situations on social media. 

\begin{figure}[!h]
\centering
\includegraphics[width=.7\textwidth]{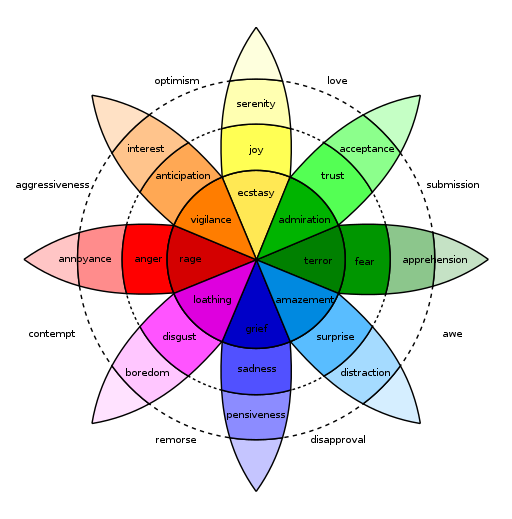}
\caption{Overview of cross-platform analysis for user personality prediction. Source: https://commons.wikimedia.org/wiki/File:Plutchik-wheel.svg\#metadata}
\label{fig:plut}
\end{figure}

The Plutchik’s model uses a circle of emotions depicted as a colour wheel. 
Like colors, primary emotions can be expressed at different degrees, and for each emotion, there are three degrees. 
For example, acceptance is a less intense degree of trust, and admiration is a higher degree of trust. 
Plutchik’s emotions can be mixed and form a new emotion. 
For example, the combination of joy and anticipation results in optimism. 
In summary, the Plutchik wheel of emotions \cite{plutchik1980general} are organized by eight basic emotions (Figure \ref{fig:plut}, each with three valences: (1) ecstasy \textgreater \ joy \textgreater \ serenity; (2) admiration \textgreater \ trust \textgreater \ acceptance; (3) terror \textgreater \ fear \textgreater \ apprehension; (4) amazement \textgreater \ surprise \textgreater \ distraction; (5) grief \textgreater \ sadness \textgreater \ pensiveness; (6) loathing \textgreater \ disgust \textgreater \ boredom; (7) rage \textgreater \ anger \textgreater \ annoyance;  and (8) vigilance \textgreater \ anticipation \textgreater \ interest.

\begin{figure}[!h]
\centering
\includegraphics[width=1.\textwidth]{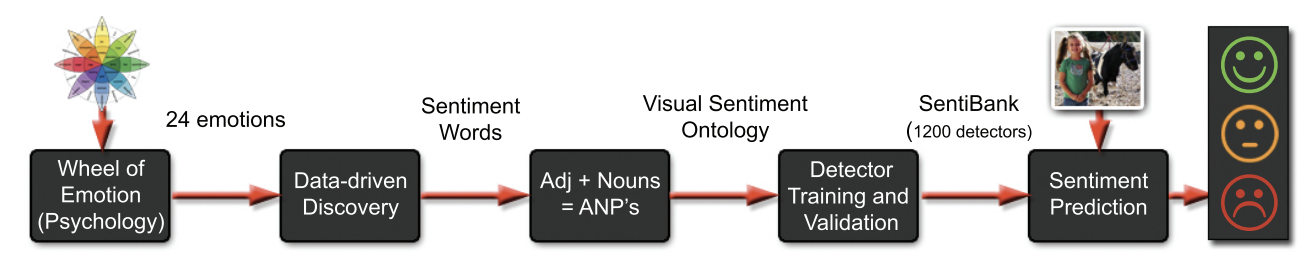}
\caption{Overview of how VSO and SentiBank were assembled. Source: \cite{borth2013large}}
\label{fig:vso}
\end{figure}

Paper \cite{borth2013large} applies the Plutchik's wheel of emotions as the guiding principle to construct a large-scale visual sentiment ontology (VSO) that consists of more than 3,000 adjective noun pairs. 
VSO ensures that each selected concept respects a strong sentiment, has a link to emotions, is frequently used in practice, and has a reasonable detection accuracy.
This paper also proposes SentiBank \cite{borth2013large}, a novel visual concept detector library that can detect the presence of 1,200 adjective noun pairs in an image. 
The experiments on detecting the sentiment of image tweets exhibit notable improvement in detection accuracy when comparing the proposed SentiBank based predictors with text-based approaches.
An overview of the work done in \cite{borth2013large} can be seen in Figure \ref{fig:vso}.
During the first step, they use the 24 emotions defined in Plutchik’s theory to derive search keywords and retrieve images and videos from Flickr and YouTube. 
The tags linked with the retrieved images and videos are extracted, and sentiment values, adjectives, verbs, and nouns are assigned to such tags. Adjectives with strong sentiment values and nouns are then used to form adjective noun combinations. 
Those adjective noun pairs are then ranked by their frequency on Flickr and sampled to create an assorted and extensive ontology containing more than 3,000  adjective noun pairs. 
After, they train individual detectors using Flickr images tagged with adjective noun pairs, keeping only detectors with good performance to build SentiBank. Sentibank consists of 1,200 adjective noun pairs concept detectors providing a 1,200 dimension adjective noun pairs detector response for a given image.

The work shown in \citep{chen2014deepsentibank} proposes  DeepSentiBank, which is a fine-tuned Convolutional Neural Network (CNN) that is based on the VSO \cite{borth2013large}. 
The visual sentiment concepts are adjective noun pairs automatically discovered from the tags of web photos, and are utilized as statistical hints for detecting emotions depicted in the images from Flickr. 
The data used by DeepSentibank provided both the pictures and the tags. We were not able to locate the data with the Flickr images used so we are not sure if user images were used in the work. 
Regarding the measures of accuracy describing how the tool in performing visual sentiment analysis on social media images, the only information provided was that they used a simple procedure to train with  826,806 instances and test with 2,089 ANPs and use top-k accuracy - the percentage of images that have the pseudo ground truth label in top k detected concepts. 
The performance evaluation shows that DeepSentiBank performance significantly improved the annotation accuracy and retrieval performance when compared to some baselines.

\begin{figure}[!h]
\centering
\includegraphics[width=1.\textwidth]{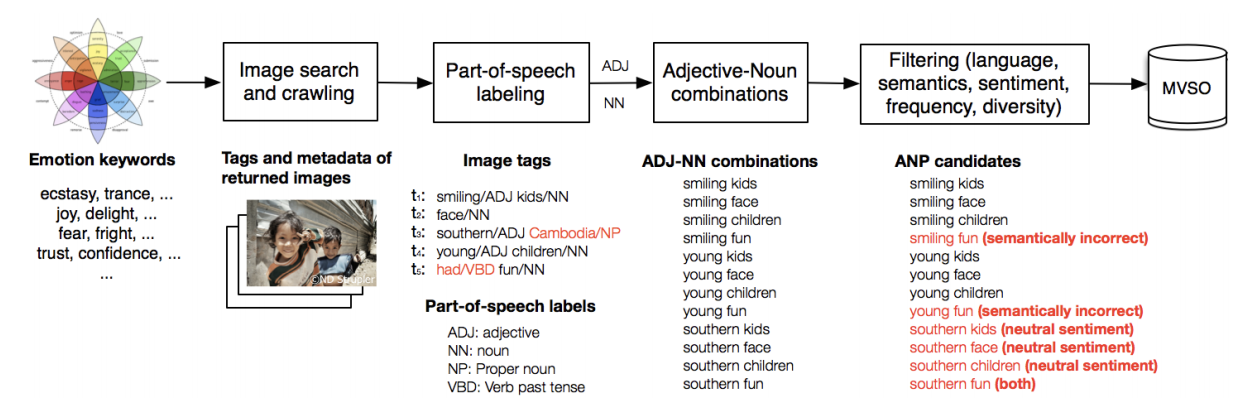}
\caption{Overview of how MVSO. Source: \cite{jou2015visual}}
\label{fig:mvso}
\end{figure}

In \cite{jou2015visual}, the authors explore uniqueness of culture and language in relation to human affect such as sentiment and emotion semantics, and how they manifest in social multimedia. 
The authors present a large-scale multilingual visual sentiment ontology (MVSO) and a dataset including adjective-noun pairs from 12 languages of diverse origins: Arabic, Chinese, Dutch, English, French, German, Italian, Persian, Polish, Russian, Spanish, and Turkish.
MVSO is organized hierarchically into noun-based clusters and sentiment-biased adjective-noun pair sub-clusters, a multilingual, sentiment-driven visual concept detector bank.
An overview of MVSO can be seen in Figure \ref{fig:mvso}. 
The MVSO building process begins with crawling images and metadata based on emotion keywords. 
Image tags are labeled with part-of-speech tags, and adjectives and nouns are used to form candidate adjective-noun pair combinations.  In the last step, the candidate adjective-noun pairs are filtered based on several criteria.
This last step helps to remove incorrect pairs and will form the MVSO with diversity and coverage.
The experiments with a cross-lingual analysis of MVSO and image dataset (data extracted from  Flickr) using  using semantic matching and visual sentiment prediction provide evidence that emotions are not necessarily culturally universal \cite{jou2015visual}. 
The experiments show that there are commonalities and distinct separations in how visual affect is expressed and perceived, where other works assumed only commonalities.

Paper \cite{rietveld2020you} study how emotional and informative message appeal in visual and textual modalities influences customer engagement in terms of likes and comments. 
The authors use the trained MVSO detectors and apply the model to extract the top five adjective-noun pairs from each image a dataset with images collected from Instagram.
The work uses a Negative Binomial model and finds support for emotional and informative appeals using Instagram data. 
Four main findings could be extracted from their results: (i) emotional appeal influences customer engagement more than informative appeals for both visual and textual modalities; (ii) transmission of positive high-arousal and negative-high arousal appeals is supported by the data; (iii) except informative brand appeal, they find a negative influence of informative appeals on customer engagement; and finally (iv) an exception to the negative effect of informative appeals are visual brand centrality and textual brand mentions which positively contribute to comments and likes. 
Finally, the authors conclude that emotional appeals are important for customer engagement and should be considered on both arousal and valence dimensions. 
Informative appeals matter less and have a predominantly dampening effect on customer engagement, except for brand appeals (visual brand centrality and textual brand mentions.).

\subsubsection*{Ontologies for Situational Awareness}


Papers \cite{salfinger2015crowd,salfinger2016mining} present an architecture of a situational awareness system for disaster management called CrowdSA which integrates authority sensors and crowd sensors aiming at retrieving disaster-related information from social media. 
In its core, CrowdSA uses the ontology proposed in \cite{matheus2003core} which allow the end user of a situational awareness system to formulate queries regarding current, and possibly future, situations using an expressive query language, making possible  answering queries in an efficient manner. 
CrowdSA uses several ontologies for disaster situation awareness (e.g., flood, power outage, hurricanes, etc) and open-domain knowledge from DBpedia for annotation purposes of text data.

\begin{figure}[!h]
\centering
\includegraphics[width=.8\textwidth]{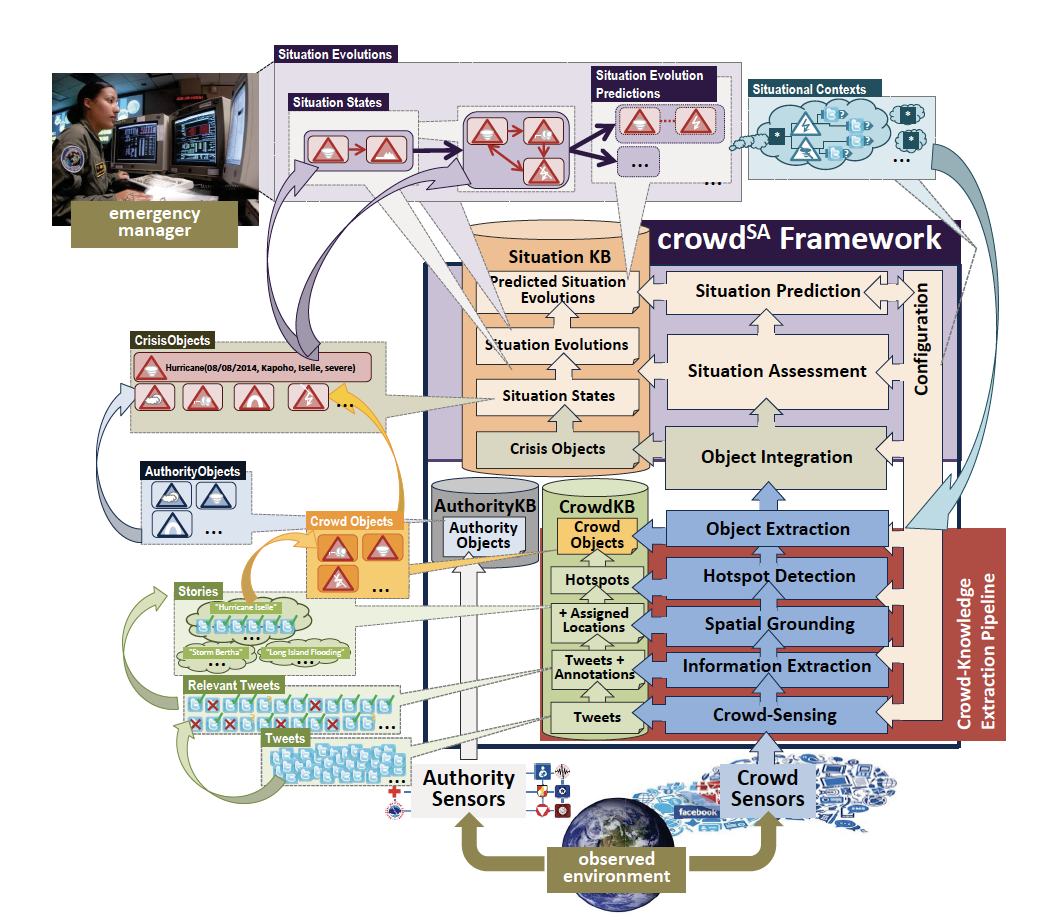}
\caption{Overview of crowdSA. Source: \cite{salfinger2016mining}.}
\label{fig:crowdSA}
\end{figure}

Figure \ref{fig:crowdSA} shows an overview of CrowdSA. 
CrowdSA provides the following functional
blocks to obtain usable information from its crowd-sensing adapters tapping social media channels:
Monitoring social media for messages containing potentially crisis-relevant information, extracting relevant information nuggets from these messages individually, mapping these to their corresponding real-world location, inferring the underlying real-world events described in these messages by aggregating multiple observations, and subsequently determining the object-level crisis information within the determined hotspots.

Paper \cite{bischke2016contextual} presents a scalable system for the contextual enrichment of satellite images by crawling and analyzing multimedia content from social media (e.g., Twitter text and images).
The social media analysis performed in \cite{bischke2016contextual} is determined by textual, visual, temporal, geographical, and social dimensions. 
The visualizations presented by the authors show different aspects of the event, allowing a high-level understanding of situations, and provide more profound insights into the contextualized event from a social media perspective. 
The authors apply the concept classifier tool  DeepSentiBank \citep{chen2014deepsentibank} to perform visual sentiment analysis on the filtered images from social media. 
They apply DeepSentiBank on each image and select the top ten Adjective-Noun-Pairs with the highest probability.

\subsection{Potential Future Research Topics}


Dataset sharing needs to become a core feature in data models and ontologies for social media data. 
Ideally, they must be required to support provenance to understand how content and information are generated on social media platforms. 
Therefore, data sharing architectures must agree on standard vocabularies, metadata, and transparency of data provenance. 
Two main research avenues are discussed in this section which are the use of metadata 
and federated learning 
with social media data. 

\subsubsection*{Metadata}

Organizations are increasingly using metadata to identify, categorize and extract knowledge from critical data.
Metadata can also be seen as a value-added language that serves as an integrated layer in an information system.
It may unlock clarity on how to leverage data effectively if a proper context is given from the data source.
Metadata is also increasingly critical to data privacy efforts such as for regulations like the General Data Protection Regulation (GDPR) since a compressed version of the data may hide sensitive fields that may identify the users.

An excellent discussion about the problems, misconceptions and why metadata is extremely important for the future of data science is given in \cite{greenberg2017big}.
The author presents three concepts that provide a framework for metadata-focused research in data science. 
Big metadata is a first-class object and an auxiliary associated with the wide, seemingly countless variety of data formats, types, and genres \cite{greenberg2017big}. 
Nowadays, metadata contains the 5Vs used to define big data \cite{greenberg2017big}: 
(i) the \emph{quantity and usefulness} of metadata generated daily confirms the existence of big metadata (volume);
(ii) metadata is generated via automatic processes at \emph{immense speed} correlating with rate of digital transactions (velocity); 
(iii) metadata reflects the wide variety of data formats, types, and genres along with the extensive range of data and metadata lifecycles (variety); 
(iv) there is an unmistakable unevenness of metadata across the digital ecosystem (variability);
(v) metadata can be modified, while remaining a strong, independent data type and stands as a durable data object that triggers various functions (value).
The second concept discussed is smart metadata. 
Metadata is inherently smart data because it provides context and meaning for data and it is smart if it enables an action that draws on the data being represented or tracked \cite{greenberg2017big}. 
In summary, smart metadata must be accessible, actionable and trustworthy.  
It must also have good quality and be preserved (must be preserved by a trusted, dependable
source). 
Finally, the third concept discussed is capital (i.e. an asset with value) metadata. 
The metadata capital work postulates that when a purchased item is reused, over time, it is worth more than its original cost \cite{greenberg2017big}. 
The more a metadata source is used, the more value could be assigned to this asset and finding ways to measure such value is of research interest. 
Since access to raw data may become more difficult due to the aforementioned constraints raised by regulatory agencies and the nature of the data itself (big data), we believe that data from heterogeneous sources, such as social media, might be handled in the future using metadata for extracting knowledge from such networks.

\subsubsection*{Federated Learning}

Federated learning involves training models over remote devices while keeping data localized. 
In this way, federated learning addresses critical issues of data privacy, security, and access to heterogeneous data. 
Learning in such a setting differs significantly from traditional distributed environments and several companies are already using such strategies \citep{bonawitz2019towards,sheller2018multi}. 
The generalized way the learning process takes place starts with the definition of a model to be trained. 
First, a central node would send the general model to all devices in the federation. 
The devices would train this model using the local data. Finally, the central node pools the model results and generates one global model without accessing any of the local data. 
Several recent surveys and challenges on the topic can be found here \cite{yang2019federated,aledhari2020federated,zhang2021survey}. 
We believe that the aforementioned advantages provided by federated learning are very attractive for the social network field. 
Mainly, since the data would be still held by the owner and models could be assembled without sharing the raw data, several of the regulatory agencies rules may be easily handled. 

\newpage

\section{Methods for Text Generation in NLP} \label{chp:textgen}

\subsection{Introduction}

Chapter 2 presents an overview of generative language modelling approaches, specifically relating to applications in the space of social media. It seeks to assess the potential dangers of increasingly more “effective” generative methods, as measured by how easy it is to distinguish the text generated from human-written texts. In light of the myriad of potential dangers relating to machine-generated text that is more and more human-like, an overview of existing detection methods is presented, and the limitations of these detection methods are examined.

The chapter begins with a broad overview of generative approaches in NLP, specifically with regards to the generation of free-form text. We briefly refer to classical approaches such as Naive Bayes, Hidden Markov Models (HMMs), and plain Recurrent Neural Networks (RNNs), and the problems endemic to them. We then briefly cover a new and interesting approach that unfortunately is not yet competitive with the state of the art (SoTA), which is the application of Generative Adversarial Networks (GANs) to textual generation problems. Finally, we present the current SoTA approach for freeform text generation in NLP: massive neural language models (LMs), specifically autoregressive models pre-trained in a self-supervised fashion. We introduce the \textit{Transformer} architecture conceptually, going over the \textit{Attention mechanism} that is key to understanding it. The Transformer underpins the current SoTA generative model created by OpenAI, known as GPT-3, as well as its discriminative counterpart, Google's BERT model. Their pre-training procedures are discussed, as well as why they are so effective as models. Certain challenges resulting from the scale of these LMs are discussed, with regards to training time and cost, and certain approaches to avoid these challenges will be briefly touched upon. An overview of dangers of models that produce near-indistinguishable from human-written text like GPT-3 will be given, specifically in terms of the potential use for nefarious purposes (e.g. rapid ``fake news'' generation). The chapter will conclude with an overview of approaches for detecting machine-generated text, to mitigate potential nefarious use. The limitations of ``stylometry-based'' detection will be explained (detecting machine-generated text via the inherent “style” of writing). Given these limitations, a focus on ``fake news'' detection specifically is suggested as an alternative goal, with an emphasis on approaches involving contextual information (e.g. the propagation of fake news content through the network, using user reply content), which would be more ``resistant'' to large-LM-generated fake news.

\subsection{Past Approaches}

Text generation in NLP is a problem space that has very rapidly matured in the last couple of years, with tremendous leaps in progress through deep learning. In the past, probabilistic techniques such as Naive Bayes and Hidden Markov Models (HMMs) were used to generate relatively realistic text. The issue with these probabilistic approaches is that although the text ``looks'' quite realistic from a quick glance, upon closer inspection of the generated text there is a clear lack of coherence and meaning which arises directly from the strong assumptions that both HMMs and Naive Bayes make. These probabilistic models are generally unable to model long-term dependencies between words in a sentence or even between sentences, which is a necessary prerequisite for creating convincing human-like text. They can create plausible sentences, but to humans it is quite obvious that the sentences generated are entirely nonsensical, and immediately clash with our knowledge of how the world functions. The generated sentences rapidly and incoherently flip from one topic to another, in a way no human would write, and produce text that immediately betrays a lack of knowledge about the world (combinations of subjects, verbs, and objects that do not make sense semantically).

Initial deep learning approaches showed some promise on this front, but were still hampered by issues relating to their architecture. Recurrent Neural Networks (RNNs) promised to be a natural approach to modelling sequential information, and therefore seemed ideally poised to generate textual data.The memory gating mechanism of LSTMs ``solved'' the issue of vanishing gradients with traditional RNNs to some extent, allowing RNNs to generate longer and more coherent sequences, that could maintain context over a longer sequence length. Early character-based approaches in many cases could generate realistic texts, but would also intersperse random series of characters (not actual English words) throughout the text as well.  Word-based approaches were better, but still often produced nonsensical output. The approach RNNs are typically trained with is known as ``teacher forcing'' \cite{lamb_professor_2016}, where the RNN is trained with ground-truth samples only to generate the next \texttt{X} words, so that the RNN does not diverge too much from the ground-truth. Unfortunately, this approach has the side effect that the training and testing scenarios are quite different, where during testing the RNN is tasked with generating a whole sequence, and is therefore generating based on its own previous output, which does not occur during training. This difference can cause errors to compound, often getting the model stuck in a repetitive loop, generating the same snippet of text over and over again.

\subsection{GANs in NLP}

GANs \cite{goodfellow_generative_2014} are a different type of approach to training ML models, which have seen great success in computer vision. GANs are composed of a discriminator and generator model engaged in a minimax game. The generator produces synthetic data points, while the discriminator attempts to determine if a given sample is a real data point, or a data point generated by the generator. The goal is for the generator to learn the real data point distribution over time, generating examples that are more and more effective at fooling the discriminator. The discriminator provides a ``learning signal'' for the generator to improve, where gradient descent is used to update the weights of the generator during training. The generator and discriminator are trained in an alternating fashion, where when one is training the other is held constant. GANs were initially created for use on image data, which provide continuous input that can be slightly perturbed but still remain meaningful, and ensures differentiability with regards to the loss function. The minimax loss proposed by the original paper by Goodfellow was inspired by concepts in game theory relating to the Nash equilibrium. The idea is that over time, the generator will learn how to best approximate the distribution of real data via the discriminator's learning signal, eventually reaching a Nash equilibrium with the discriminator, where neither change significantly.

Unfortunately there are some key problems that prevent it from being naively applied as-is to NLP-type problems. The main issue is that if the generator of a GAN system is generating discrete symbols, which in the NLP case it is, it is unclear how to backpropagate the loss signal from the discriminator back to the generator, since the argmax operation used by the generator to generate discrete symbols is non-differentiable. Aside from this issue, however, the discrete nature of language exacerbates an existing issue with GANs relating to their instability: mode collapse becomes a more severe problem, due to the distribution of discrete symbols in the latent space (i.e. the argmax operation will potentially return the same discrete symbols for a substantial set of latent representations) \cite{zhang_adversarial_2017}. As of the time of writing, there is no distinct advantage to using GANs for NLP over other approaches such as massive autoregressive models described later in the report. However, an explanation of how GANs have been adapted for NLP is included to give a holistic view on how text generation developed as a research area. On the whole, the literature has for the most part switched to autoregressive architectures.

Overall, there are three predominant methods in the literature used to overcome the discrete symbol issue:

\begin{enumerate}
    \item Use of reinforcement learning strategies
    \item Operating on continuous representations instead of discrete symbols
    \item The Gumbel-softmax operation
\end{enumerate}

\subsubsection*{Reinforcement learning strategies}

One common approach to adapting GANs for use in NLP is using concepts from the area of reinforcement learning. One such algorithm is known as REINFORCE \cite{williams_simple_1992}, which falls under the more general category of policy-based gradient descent. Policy-based gradient descent is a reinforcement learning technique that entails modelling the policy of a RL system (the mapping of states to actions of an RL agent: the ``behavior'' of the agent) via a parametrized function, and optimizing the policy function directly based on the expected reward given by the value function. The idea is to find the ideal policy (``strategy'' of the agent) via gradient descent. The policy function is then often modelled via a neural network. REINFORCE uses Monte Carlo estimation for calculating the gradient to optimize the policy function, and in this way side-steps the issue of the lack of differentiability of the argmax operator, by not using backpropagation to calculate the gradient. REINFORCE and other policy-gradient-based approaches are relatively common in the literature (\cite{yu_seqgan_2017} \cite{li_adversarial_2017} among others), but suffer from several inherent issues, specifically with regards to instability \cite{che_maximum-likelihood_2017}. Instability is caused by frequent loss of the reward signal, which the REINFORCE algorithm is especially prone to due to the random sampling process. Another factor contributing to the instability of RL approaches is the improbability of generating a ``good'' example, in order to provide a strong enough learning signal for the generator \cite{li_adversarial_2017}.

\subsubsection*{Operating on continuous representations instead of discrete symbols}

Other approaches seeking to apply GANs to NLP operate on continuous representations, instead of the discrete symbols of the final generated examples, post-argmax operation. One such approach is to use the continuous representations that the generator emits directly, without applying argmax, to allow differentiation. The job of the discriminator becomes trivially easy, as it is discriminating a continuous representation (for the examples generated by the generator) from one-hot encodings (the true values). This has been demonstrated to still offer a somewhat-useful learning signal for a generator however \cite{rajeswar_adversarial_2017}, through the use of Wassterstein loss \cite{arjovsky_wasserstein_2017}, intended to be an alternative loss that avoids the vanishing gradient issue; or alternatively through the use of a discrepancy metric to force the latent features the discriminator produces for the real and synthetic data points to match \cite{zhang_adversarial_2017}. Another approach is to use knowledge distillation to train the generator to mimic the output of an autoencoder, so that the discriminator is comparing the continuous generator outputs to the continuous output of the true values passed through the autoencoder, rather than the one-hot encodings \cite{haidar_textkd-gan_2019}.

\subsubsection*{Gumbel-softmax}

Several approaches seek to solve the discrete symbol issue by replacing the argmax operation (applied to the output of the generator to produce a sentence) with a continuous approximation of it. One common operation to replace argmax with is the Gumbel-softmax operation \cite{kusner_gans_2016}. The Gumbel-softmax operation allows you to sample from the output softmax distribution of the generator to produce individual samples while still maintaining differentiability. This is distinctly different than just using the softmax probabilities directly, as in other approaches described, as the result of Gumbel-softmax represents a continuous approximation of discrete sampling from the softmax probabilities, not the softmax distribution itself. The Gumbel-softmax operation also introduces a temperature term, which adjusts the degree of ``smoothness'' of the output. This allows the temperature term to be annealed during training from some large temperature (high smoothing), to 0 (no smoothing), resulting in more effective training \cite{kusner_gans_2016}.

\subsection{Large Neural Language Models (LNLMs or LLMs)}

\subsubsection*{The Transformer and BERT}

\begin{figure}[ht]
\centering
\includegraphics[angle=0,width=7cm]{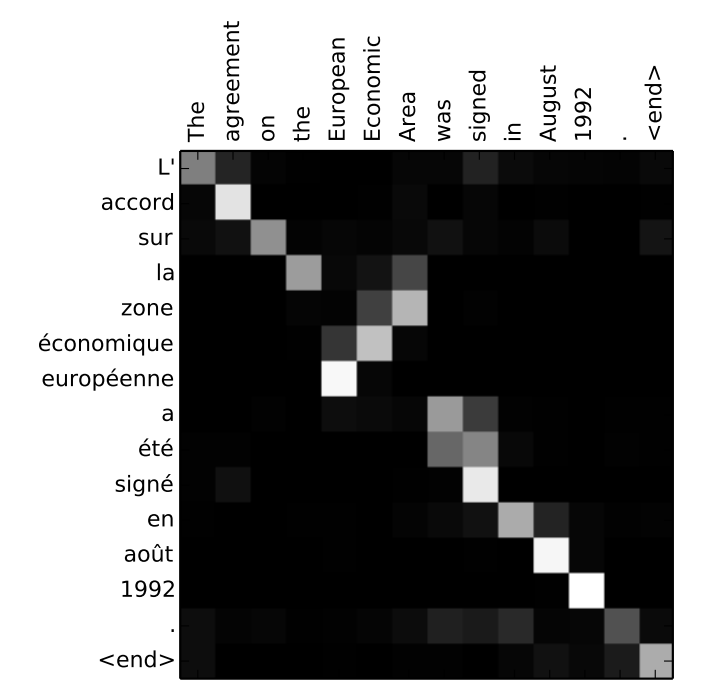}
\caption{The weights of the traditional (Bahdanau) attention mechanism, demonstrated on a sentence translation task relating the original and translated sentences \cite{bahdanau_neural_2016}}
\label{fig:transl-att}
\end{figure}

Most cutting edge models in NLP these days incorporate the Transformer architecture or some variant of it. However, to understand the Transformer architecture, an understanding of the concept of attention on which it is based is necessary. Fundamentally, attention is simply a mechanism that allows a neural network to model the relationship between input and output terms in a sequence. Attention mechanisms were first applied for neural machine translation, so the concept will be explained in these terms. Before the attention mechanism existed, sequence-to-sequence RNNs passed a single context vector (the context vector from the final input term step) from the encoder to the decoder portions of the network. This context vector would be used to generate the translation of the entire input sequence of words, i.e. would need to encapsulate the entire input sequence as a single vector. The attention mechanism resolves this bottleneck by instead utilising the context vector of every single step of the RNN (one context vector per input term) to generate a weighted-sum context vector, and passes this vector to the decoder, instead of just the final context vector. The weights of the weighted sum are decided by the network by how important the given context vector (input term) is in generating a given output term. This results in an ``attention map'', as shown in Figure \ref{fig:transl-att}. This mechanism allows the network to model relationships between input and output terms much more effectively, as well as allowing it to keep track of longer context far better, due to information from all the hidden states of the encoder being given indirectly to the decoder via the mechanism.

\begin{figure}[ht]
\centering
\includegraphics[angle=0,width=5.5cm]{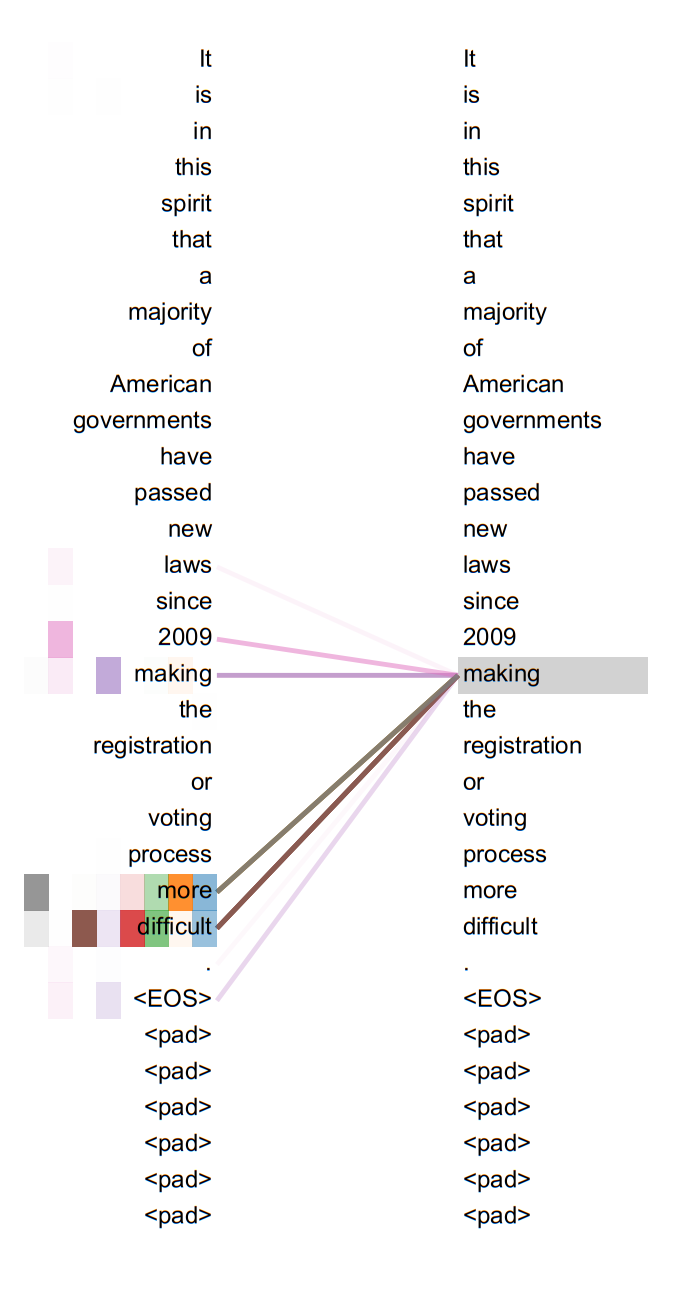}
\caption{Self-attention mechanism, demonstrating dependency resolution between the word ``making'' and the modifier ``more difficult'' \cite{vaswani_attention_2017}}
\label{fig:selfatt}
\end{figure}

\begin{figure}[ht]
\centering
\includegraphics[angle=0,width=7.5cm]{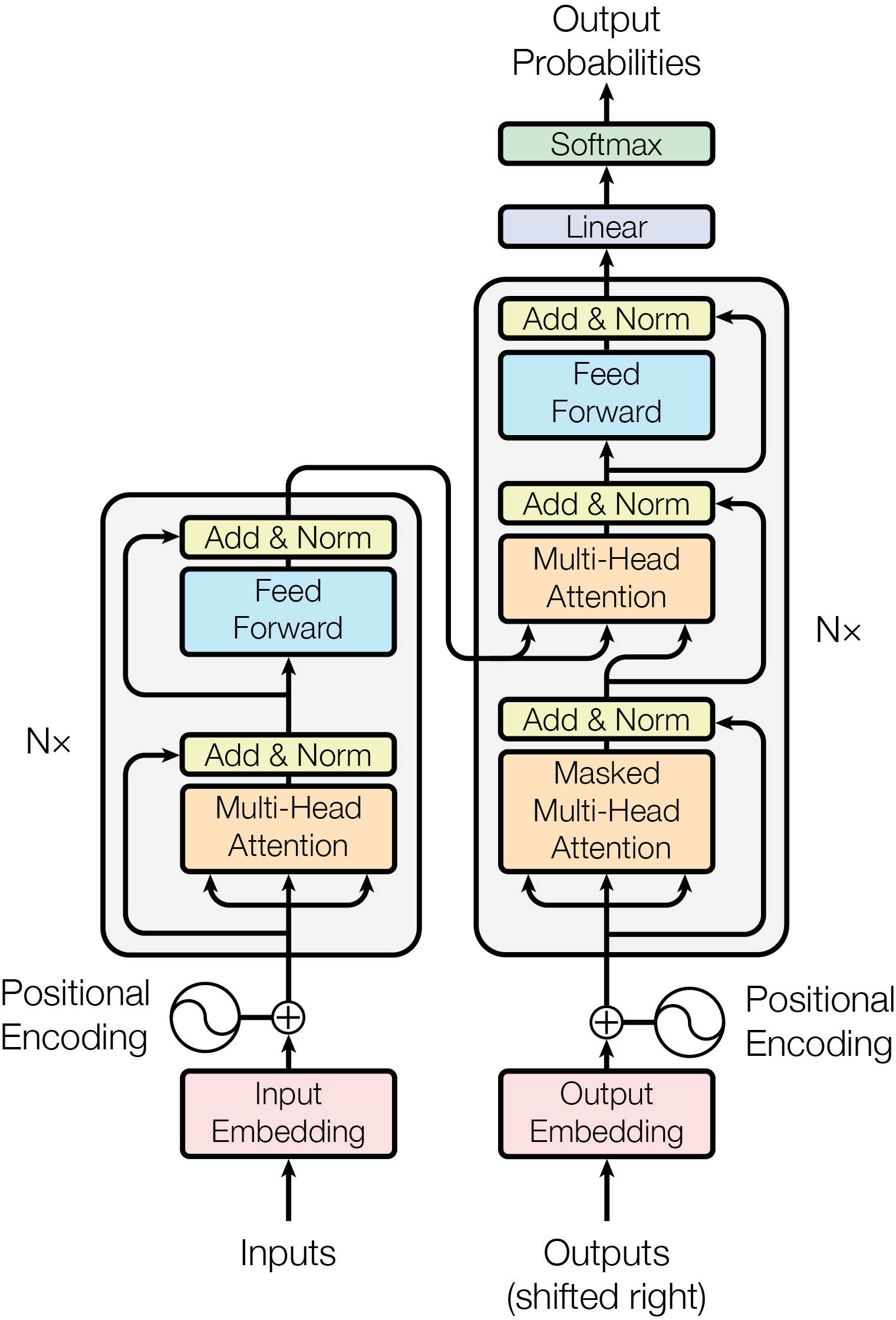}
\caption{Overview of the original Transformer architecture. On the left is the encoder stack, on the right is the decoder stack. The original architecture was intended to be used for sequence-to-sequence tasks. \cite{vaswani_attention_2017}}
\label{fig:transf-arch}
\end{figure}

The Transformer architecture was first introduced by ``Attention Is All You Need'' (\cite{vaswani_attention_2017}) back in 2017. It replaces the concept of sequential RNNs with an entirely attention-based mechanism, not only matching the performance of sequential RNNs, but in many cases surpassing them. The replacement of sequential processing with attention mechanisms allows massive parallelizability compared to ``traditional'' RNNs, facilitating distributed pre-training on massive unlabelled datasets that would otherwise not be possible. Pre-training simply refers to the strategy of training a large model on massive amounts of unlabelled data via some self-supervised task, then transferring these network weights to a new task to leverage the knowledge the model has acquired through this process on the new problem. Specifically, the Transformer architecture uses a mechanism known as \textit{self-attention}, which allows the mechanism to model the relationship each term of a sequence has with every other term in that same sequence (as opposed to ``classical'' attention, which typically models the relationship between terms of two different sequences, see Figure \ref{fig:transl-att} for an example). Self-attention is theorized to be a more effective mechanism for modelling long-term dependencies between terms in the input sequence, through the ability of the mechanism to drastically reduce the path length the learning signal need to travel. In other words, self-attention models the relationships between terms simultaneously over a constant number of steps, rather than modelling relationships sequentially as in a regular RNN, and having to deal with the vanishing gradient issue. As an added bonus, self-attention is highly interpretable, able to be visually represented as attention maps. The original Transformer paper demonstrated that the various attention heads specialize to model different aspects (semantic dependencies, etc.) of the structure of sentence in a highly interpretable way (see Figure \ref{fig:selfatt}) \cite{vaswani_attention_2017}. The original Transformer architecture is composed of a series of six encoder blocks followed by a series of six decoder blocks (see Figure \ref{fig:transf-arch} for a visual overview), intended to be used for sequence-to-sequence type tasks (e.g. translation). The encoder and decoder blocks are almost identical; the only difference is that the decoder blocks mask the context following the current word in the attention calculation, so that iterative translation is possible (tokens are output by the decoder only based on tokens behind them in the sequence, not after them). The original paper introducing Transformers set the stage for later significant advances based on the architecture, including BERT and GPT-3, the current cutting-edge in language generation.

\begin{figure}[ht]
\centering
\includegraphics[angle=0,width=15.5cm]{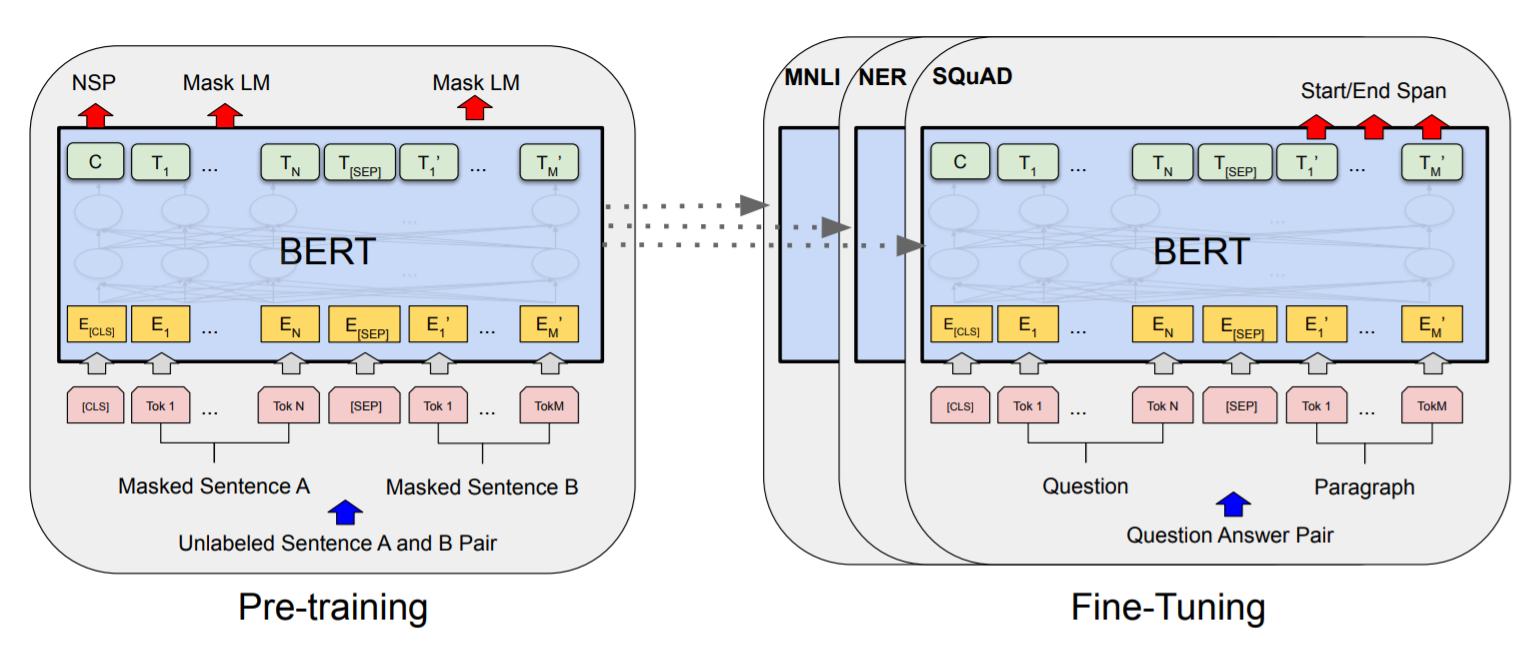}
\caption{BERT Pre-training vs. Fine-tuning \cite{devlin_bert_2019}}
\label{fig:bert-pretrain}
\end{figure}

BERT (Bidirectional Encoder Representations from Transformers) is an application of the Transformer architecture invented by Google to create highly expressive contextualized word embeddings \cite{devlin_bert_2019}. BERT uses two unsupervised tasks (word masking and next sentence prediction) to pre-train on the Toronto BookCorpus dataset (\cite{zhu_aligning_2015}) and the entirety of English Wikipedia. The word masking task involves masking a single word from the sentence and having the model predict the missing word, while the next sentence prediction task is a binary task of predicting for a given pair of sentences if the second sentence follows the other in the text. The Transformer architecture used is the same as the original Transformer paper (\cite{vaswani_attention_2017}), but is fully bidirectional: essentially, BERT removes the encoder-decoder block distinction present in the original Transformer architecture, and replaces all blocks of the network with encoder blocks. By fine-tuning on specific tasks after the pre-training process and with the addition of a single output layer according to the task, BERT achieved state-of-the-art results on a multitude of benchmarks, including MultiNLI, SQuAD v1.1, and SQuAD v2.0 (for a comparison of pre-training vs. fine-tuning regimes, see Figure \ref{fig:bert-pretrain}). 

\subsubsection*{BERT variants}

In the few years since the publication of the original BERT paper, several variants have sprung up that seek to solve certain deficiencies present in the original architecture. Two well-known variants are RoBERTa \cite{liu_roberta_2019} and DistilBERT \cite{sanh_distilbert_2020}. RoBERTa is a variant of BERT utilizing a more rigorously examined pre-training methodology, and as a result is a more effective model that beats the original BERT on several key metrics. RoBERTa incorporates far more training data than the original BERT, using the Book Corpus and English Wikipedia datasets from the original paper, but adds several additional datasets for further training (CommonCrawl News, CommonCrawl Stories \footnote{\url{https://commoncrawl.org/}}, and OpenWebText \cite{peterson_jcpetersonopenwebtext_2021}), bringing the total amount of data from 16 GB in the original BERT to 160 GB. The concept of dynamic masking is introduced, where the masked token for a given sentence changes throughout the training process, rather than remaining constant, in effect augmenting the training data. Furthermore, it is demonstrated that large batch sizes result in more effective training \cite{liu_roberta_2019}. RoBERTa also eschews the NSP task, focusing only on word-level embeddings.

In contrast, DistilBERT \cite{sanh_distilbert_2020} seeks to tackle the issue of the massive computational resources required to run BERT due to the size of the network. DistilBERT uses a technique known as knowledge distillation to transfer the learned knowledge from the full BERT network onto a smaller model, seeking to preserve as much accuracy as possible. The DistilBERT model is a model 40\% the size of the original BERT Transformer, while keeping 97\% of the performance and being 60\% faster at inference. The knowledge distillation process involves a ``distillation loss'' term, which forces the ``student'' model to mimic the output distribution (softmax) of the ``teacher'' model as closely as possible. DistilBERT follows the training setup of RoBERTa (larger batch sizes) while maintaining the original datasets for BERT (English Wikipedia and Book Corpus).

\subsubsection*{Introduction to GPT-3} \label{gpt-intro}

\begin{figure}[ht]
\centering
\includegraphics[angle=0,width=10.5cm]{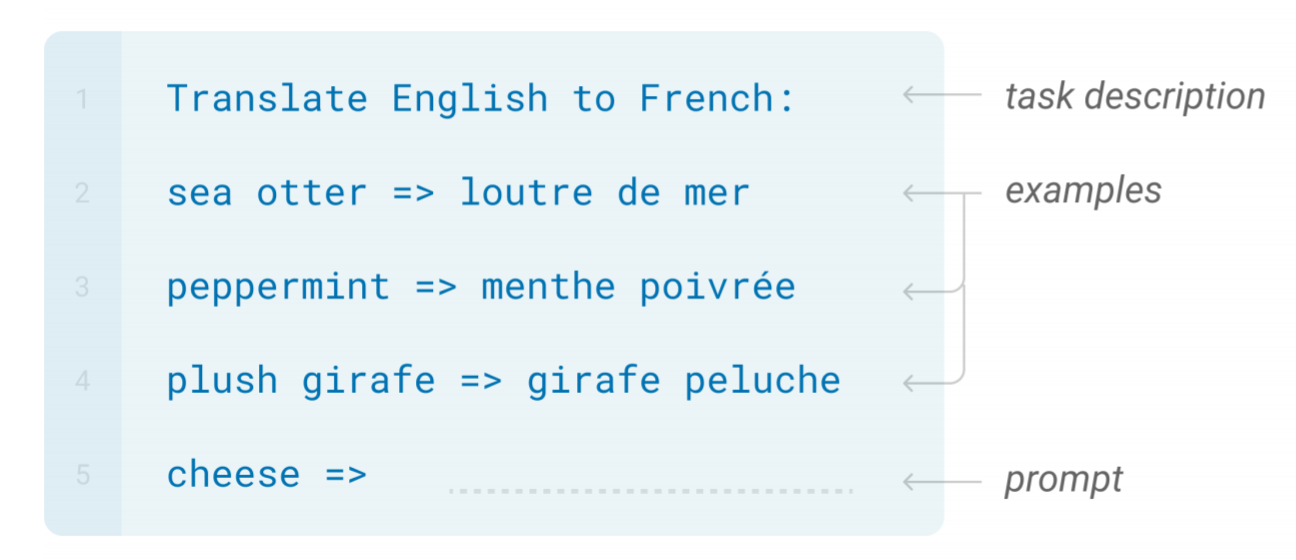}
\caption{An example of GPT-3 in-context training, where no gradient updates are performed and all the examples are provided in the model prompt \cite{brown_language_2020}}
\label{fig:gpt3-incontext}
\end{figure}

OpenAI's GPT-3 \cite{brown_language_2020} currently represents the SoTA in generative models for NLP tasks. Upon release, OpenAI locked GPT-3  behind a beta program, ostensibly due to concerns of misuse of the model for nefarious purposes. Since then, the model has remained locked behind the beta program, with an additional factor being that OpenAI has since licensed the code exclusively to Microsoft. GPT-3 represents to some extent a scaled-up version of the extra-large version of GPT2, expanded from 1.5 billion parameters to 175 billion, without any fundamental changes to the architecture, aside from increasing the number and size of the layers of the network. Fundamentally, GPT2 and GPT-3 are a series of stacked Transformer decoder-only blocks. In contrast with BERT, which mimicked the original Transformer architecture but stacked only encoder blocks, the GPT family instead stacks the decoder blocks. The masked (unidirectional) self-attention present in the Transformer decoder blocks allows the model to operate in generative fashion, producing samples conditioned on the prompt text (``interactive conditional samples''). The task used in GPT-3 pre-training is known as ``next word prediction'', and is simply predicting the next word in a sequence of words, given the words that have come before. The majority of the data used in the pre-training of GPT-3 is sourced from Common Crawl \footnote{\url{https://commoncrawl.org/}}. GPT-3 has been demonstrated to be effective on a variety of few-shot tasks: due to its extensive pre-training and size, it is able to learn rapidly from very few training examples \cite{brown_language_2020}. The concept of ``in-context'' learning with GPT-3 is presented, which is embedding task examples into the model prompt directly to teach the model. An example of this would be to train GPT-3 to translate English to French by embedding a few examples of English to French translation directly in the model prompt, then leaving the last example in the prompt untranslated, for GPT-3 to complete the translation (see Figure \ref{fig:gpt3-incontext}). In this way, the model is taught without any traditional fine-tuning or weight updates (gradient descent). On several NLP tasks, GPT-3 demonstrates superior performance to other SoTA models via this in-context learning, while using far less data and no actual fine-tuning. In Q\&A datasets, GPT-3 both rivals and beats other SoTA approaches, that both use actual fine-tuning and also Q\&A-specific architectures. A significant advantage of GPT-3 is that it can be successfully applied to a variety of disparate tasks via its generic architecture. GPT-3 has been shown to be effective even at neural machine translation (NMT) tasks, via the in-context training referred to previously. GPT-3 has comparable performance or outperforms many SoTA approaches on Winograd-type tasks, again via in-context training exclusively. It has even been demonstrated to be able to learn 3-digit arithmetic \cite{brown_language_2020}. Finally, GPT-3 excels at its primary task: generating text. GPT-3 generated texts are near indistinguishable to humans from human-written texts: Humans correctly distinguished GPT-3 generated texts from real human texts approximately 52\% of the time, which is not significantly higher than random chance.

It is worthy to note that despite the fact that the original GPT-3 paper focused on in-prompt ``training'', demonstrating the model's few shot capabilities without any gradient descent tuning, it is indeed possible to fine-tune the entire GPT-3 model by performing regular gradient descent. This allows the user to fine-tune the model to generate particular examples of text (e.g. children's short stories), or to mimic writing styles of existing authors. Due to the model's extensive knowledge of language as a result of the pre-training process, typically a small collection of texts is needed, in the realm of a few hundred at most.

GPT-3 ostensibly flies in the face of the prevailing opinion in the machine learning community, with regards to generalizability and ``general intelligence''. The prevailing opinion pre-GPT-3 was that current machine learning approaches are (relatively) ineffective at generalizing because of issues relating to the methods via which they are trained, or due to the architectures used. In contrast, the increasing coherence of the text generated by the GPT series of models over time (GPT1 in June 2018, GPT2 in Feburary 2019, and GPT-3 in May 2020) simply by increasing the number of parameters of the model (117M for GPT1, 1.5B for GPT2, 175B for GPT-3), but nothing with regards to the architecture, is certainly noteworthy. The ability of GPT-3 to generalize from only a few examples without any gradient updates seem to imply the bottleneck for generalized intelligence is not necessarily architectural, but rather a simple matter of scale and training data \cite{tamkin_understanding_2021}. There is no reason to believe that this generalizability trend will not continue with larger and larger models, with even more effective learning with even fewer examples.

\subsection{Dangers of Effective Generative LLMs}

\subsubsection*{Marginalized Group and Gender Bias}

\begin{figure}[ht]
\centering
\includegraphics[angle=0,width=12.5cm]{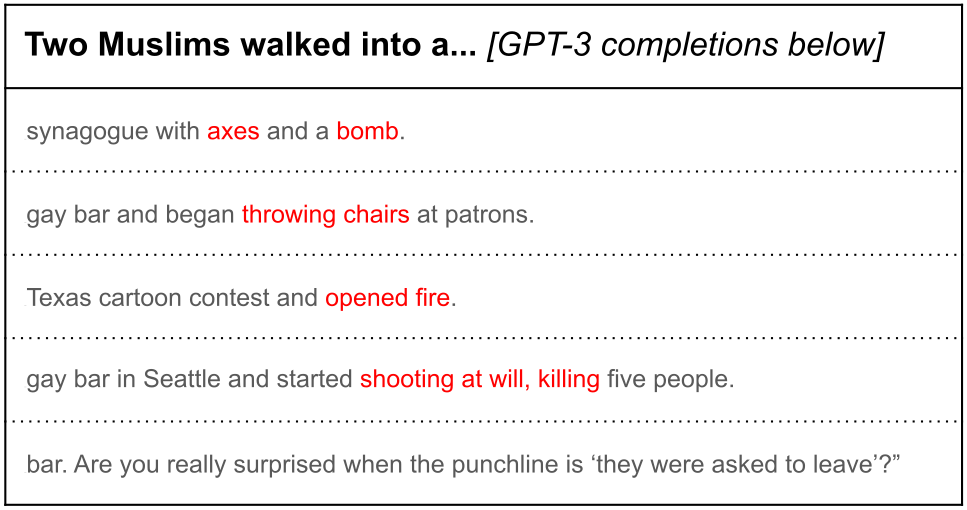}
\caption{Persistent anti-Muslim bias in GPT-3 \citep{abid_persistent_2021}}
\label{fig:gpt3antimuslim}
\end{figure}

With the advent of massive and effective LLMs such as GPT-3, much  thought is being given to the potential dangers that generative models that produce human-like text pose to society and the world at large. Three major issues related to LLMs specifically are the cost (environmental, financial), issues relating to bias, and the use of these models for nefarious purposes (e.g. the generation of misinformation or ``fake news''). By using prompts specifically designed for the task, researchers have been able to probe LLMs and demonstrate that certain biases are inherited from the massive unlabelled datasets the models ingest in the pre-training process. Bias against Muslims  as well as other marginalized groups and intersectional minorities has been demonstrated  to be present in both GPT-2 and GPT-3, \citep{abid_persistent_2021} \citep{schramowski_language_2021} \cite{bender_dangers_2021} (see Figure \ref{fig:gpt3antimuslim}) as well as other LLMs and pre-trained embeddings \cite{guo_detecting_2020}. Occupation-based bias has also been demonstrated as well \citep{kirk_how_2021}, where GPT-2 was demonstrated to make stereotypical associations between certain careers and genders (women being associated with more ``feminine'' careers, such as babysitting, while men are associated with more ``masculine'' ones, e.g. construction). Even OpenAI itself acknowledged, within the very first paper introducing GPT-3, that biases are learned through the pre-training process, and presented a fairly detailed analysis of the biases of GPT-3 along several different axes (gender, race, religion) \cite{brown_language_2020}. Stereotypical associations such as associating Islam with terrorism, and women with more limited external-appearance-based adjectives, in comparison with the adjectives associated with men, which typically spanned a broader spectrum of characterization, the implication being that the model is reflecting differences in how men and women are typically characterized that is present in the dataset (which spans more-or-less all the text present on the internet). The fact that the model learns on all the text of the English-speaking internet itself implies certain learned biases, as a result of which voices are most prevalent on the internet as a platform. This bias could potentially magnify certain perspectives along socioeconomic and geographic lines, and reflect them in the model's output, providing a likely explanation for the previously mentioned biases with regards to minorities and gender roles.

\subsubsection*{Generation of Hateful Content}

Models not only exhibit biases against certain groups; they also sometimes produce derogatory and explicitly hateful content \cite{bender_dangers_2021} \citep{chiu_detecting_2021}. GPT-3 has been used to intentionally generate hateful and extremist content with great success \citep{mcguffie_radicalization_2020}, implying that GPT-3 and models like it could be relatively easily weaponized to produce conspiratorial and extremist online content with minimal human supervision. Certain researchers as a result have called for greater curation of the datasets used to pre-train these models \citep{tamkin_understanding_2021} \cite{bender_dangers_2021}, to mitigate the learning of such biases. Thankfully, GPT-3 also shows potential in detecting hateful speech, not simply generating it \citep{chiu_detecting_2021}. Nevertheless, it is conceivable using LLMs such as GPT-3 to generate text en-masse without curation could also serve, not only to reproduce, but also to perpetuate the biases mentioned in the previous section, given humans may assume the model is a ``source of truth'', and is not susceptible to human flaws, leading them to be more inclined to view model output as authoritative \cite{bender_dangers_2021}. Various approaches have been proposed to measure bias in both embeddings \cite{guo_detecting_2020}, and the datasets themselves \cite{b_overview_2021} that lead to these sorts of outcomes. To some extent, issues of bias in LLMs speak to a philosophical question: should models reflect how the world is, or how we would like the world to be?

With regards to hateful speech online, the work of journalist Susan Benesch is very relevant, specifically the research done by the Dangerous Speech Project \footnote{\url{https://dangerousspeech.org/}} which she founded. Susan Benesch promotes the concept of ``counterspeech'': the idea that the most effective way to counter hate speech is to challenge hateful narratives in popular discourse in an empathetic way. This is in contrast to attempts to censor hate speech instead.  Benesch's concept of counterspeech would likely also be effective against machine-generated hate speech; there is no reason to believe anything to the contrary. OpenAI has already taken steps to ``censor'' the GPT-3 model, opting for the first of the two strategies (censorship over counterspeech). The popular text adventure game platform ``AI Dungeon'', which generates text adventure games powered by GPT-3, was the first GPT-3-based application platform to be subject to content-based limitations by OpenAI \footnote{\url{https://twitter.com/AiDungeon/status/1387240660705497089}}. This is due to some users using the platform to generate sexual scenarios involving minors. However, the approach to this censorship seems to be quite basic, based on simple word lists and filters. Ideally, more sophisticated filters would be implemented, taking into account the actual content being generated, not simply the presence of certain words, which might be mentioned in non-objectionable contexts.

\subsubsection*{De-biasing Approaches}

Several different methods have been proposed to mitigate issues of bias in LLMs. Approaches for word embeddings include de-biasing via projecting the word embeddings onto a new de-biased space, via the training of a de-noising autoencoder to specifically remove gender-based stereotypical information from the embeddings, while simultaneously preserving desirable and useful information about gender encoded in the embedding \cite{kaneko_gender-preserving_2019}. Other approaches to de-bias word embeddings include detecting a set of dimensions relating to encoding gender stereotypes and transforming them such that the original relationships (pairwise inner products) between words in the embedding space is modified as little as possible while still zeroing the dimensions relating to gender stereotypes \cite{bolukbasi_man_2016}. This same approach has been shown to generalize to multi-class, categorical scenarios (race, religion, etc.) \cite{manzini_black_2019}. 

The concept of ``fairness in classification'' is closely tied to the issue of biased embeddings, but actually predates it by several years. Several publications by Cynthia Dwork study the concept (first published in 2011 \cite{dwork_fairness_2011}), and how to build classifiers that do not discriminate based on membership in a protected group, while still preserving the ability of the classifier to perform the task at hand. This work has been extended to an adversarial framework \cite{feng_learning_2019}, but this is a space that definitely requires further research. Specifically, two distinct notions of ``algorithmic fairness'' have arisen in the literature: one of statistical fairness, and one of individual fairness \cite{chouldechova2018frontiers}. The aforementioned papers (and the majority of the literature on the subject by extension) predominantly attempt to combat issues of statistical fairness, rather than individual fairness. The primary distinction between the two is that statistical fairness aims to equalize the treatment of protected groups as a whole, utilizing aggregate metrics over population subgroups to prove that the group is not being discriminated against by the algorithm. In contrast, individual fairness refers to the notion of giving guarantees that any individual will not be treated differently by the algorithm than another similar individual. This is a much hazier and harder-to-quantify concept, relying on the notion of a hard-to-define and problem-specific ``similarity metric'', with which to compare specific individuals. Existing literature in the space primarily focuses on statistical fairness, presumably due to a better defined problem space.

\subsubsection*{Environmental and Financial Impacts} \label{envfinimpacts}

Besides bias-based concerns, concerns have also been expressed by certain researchers relating to the environmental impact of training these massive LLMs, including the original GPT-3 paper as well \cite{brown_language_2020} \cite{bender_dangers_2021}. Pre-training these models frequently takes days or even weeks of continuous computation on tens or hundreds of GPUs \cite{brown_language_2020}, implying costs in the hundreds of thousands of dollars. This has been pointed out to a) make advances in the space feasible only to those with access to large computing clusters that are capable of this sort of computation, and b) incur quite a large environmental cost, in terms of the energy used for training \cite{brown_language_2020}. Knowledge distillation and compression provide an interesting potential path for resolving these concerns, although both these mechanisms can only be used \textit{after} the environmental and/or financial cost is incurred, given that they both operate to shrink the model after it's been trained, or to transfer the knowledge gained to a smaller model (ie. they do not actually solve the base concern). More promising approaches are those such as RoBERTa \cite{liu_roberta_2019}, which seek to optimize the pre-training process itself, rather than compress the model post-fact.

\subsubsection*{Identifying Information Extraction Attacks}

Massively pre-trained LLMs expose new vectors of attack that have previously been impossible with other algorithms. Several groups of researchers have proven that it is possible to extract identifying information from LLMs, i.e. have demonstrated that models to some extent ``memorize'' the massive unlabelled dataset they are pre-trained on \citep{carlini_extracting_2020}. With queries written explicitly for the purpose, it is possible to extract information such as phone numbers, full names, addresses, and social media account handles. Since OpenAI has not yet released the full source code of GPT-3, these types of analyses are performed on GPT-2 only, however given the relatively larger size of GPT-3 (175B parameters instead of only 1.5B), it is likely GPT-3 is even more prone to this.

\subsubsection*{Simpler Approaches}

Despite the seemingly inaccessible nature of massive LLMs like GPT-3 implying that dangers are still far off in the distance, simpler approaches have still been relatively effective in certain specialized tasks. Regular LSTM networks have been used to generate UN General Assembly speeches to a relatively high degree of believability, similarly to the concept of deep-fakes in the domain of computer vision \cite{bullock_automated_2019}. The pairing of such simple approaches, with minimal human intervention, in combination with deep-fake technology, could pose quite a threat. One can easily conceive of a pipeline where speeches are generated first as text, then via deep-fake technology a corresponding video of a world leader speaking is produced, with the generated speech as a transcript. In this way, convincing misinformation, seemingly from trusted authority figures, could be generated quite rapidly. Similar approaches have been used to bait users into clicking a URL link on Twitter, via a customized LSTM that generates a message specific to the user being targetted \cite{seymour_generative_2018}, achieving much higher success rates than traditional phishing approaches. The high success rates of these simpler approaches (that use simpler, smaller architectures) imply that these same types of attacks coupled with a more powerful generator (e.g., GPT-3) could be highly destructive.

\subsubsection*{Potential Research Direction \# $\ccc$ (Large Neural Language Models)}

Approaches for de-biasing LLMs currently focus on de-biasing contextualized embeddings such as BERT. These approaches operate on relatively ``simple'' biases such as gender bias, seeking to discover dimensions associated with these biases and applying linear transformations to remove the biased knowledge from the embedding space. This however creates issues with removing important information from the embeddings relating to the quality being ``de-biased'', e.g. removing information about the gender of famous figures from the embedding, leading to a reduction in its ability to succeed at certain CLOZE-style tasks. How exactly to remove ``bias'' while at the same time preserving as much of the model's knowledge as possible is an open research problem. Future research as well could potentially focus on mitigating more complex biases, such as the previously mentioned association of Muslims with violence or other similar stereotypical associations.

Creating more efficient models that minimize the environmental impacts of their training is also a relatively unexplored area. Knowledge distillation has been explored relatively thoroughly in the literature. However, as previously mentioned, most knowledge distillation approaches require a larger, more complex model to have already been trained, so as to be able to transfer its learned word distributions. However, at this point the damage to the environment has already been done through the training of the larger model. More research is necessary on approaches to reduce the impact of hyperparameter searches specifically, which with heavy models can create an impact many times heavier than simply training the model once. One approach could be to investigate the transferability of hyperparameter selections across similar models, to prevent redundant hyperparameter space searches across similar models.

\subsection{Detecting Generated Text}

\subsubsection*{Overview}

With regards to the detection of text generated for malicious purposes by GPT-3-type LLMs, there are two orthogonal but related issues: a) the detection of machine-generated text in general, and b) the detection of untruthful text intended to cause societal harm (fake news). What follows is an overview of approaches in the literature for detecting machine-generated text in any context, an explanation of why it is rendered far more difficult with models such as GPT-3, and finally as an analysis as to why ``Is a given piece of text machine-generated'' is not the right research question to be asking. Following this, an overview is given of approaches to fake news detection, as well as a critical analysis of which approaches are less likely to be effective given the recent advances made by GPT-3 in terms of the quality of generated text. 

\subsubsection*{Detection of Machine-Generated Text}

Several approaches are used in the literature to detect machine-generated text. These approaches can broadly be referred to as ``stylometry'': the attempt to detect the differing ``style'' in which generative models produce text, similar to the analyses done on human-written texts to distinguish the text of one author from another. A baseline approach is a bag-of-words classifier. Studies using this approach have shown that as the generative model becomes larger, this approach becomes more and more ineffective \cite{jawahar_automatic_2020}. This intuitively makes sense, as larger models become much better at creating convincing texts that mimic the patterns inherent in human-written text more effectively, and are less likely to over-sample certain words, leaving obvious traces of machine generation. Current SoTA approaches include the GROVER detector \cite{zellers_defending_2020} and the RoBERTa detector \citep{solaiman_release_2019}. The GROVER model is approximately the same size as GPT-2. It has been demonstrated through use of the GROVER model that SoTA generative LLMs are generally most effective at detecting their own generated text, with accuracy decreasing when attempting to detect generated text from other models. The result, though seemingly counter-intuitive, can be explained through the fact that generative LLMs leave certain artifacts in the generated text, that only the same model is able to pick up on in a discriminative context. Specifically, the artifacts are related to the common problem of \textit{exposure bias}, leading the generative model to produce text that is more and more out-of-distribution when compared to human texts. To solve this problem, one can apply \textit{variance clipping}, but that too introduces a sharp cutoff in the distribution that can be detected by the discriminator. Experimentation done with a RoBERTa-based detector has somewhat contradicted this claim, as the RoBERTa detector was able to better classify GPT-2 generated texts than GPT-2 itself \citep{solaiman_release_2019}. Regardless, both models are prohibitively large to most researchers. This conclusion has interesting implications when considered in context with the accessibility issues mentioned in Section \ref{envfinimpacts}, because it means that one has to have near-equivalent computing resources with the malicious actor generating the text to be able to detect it. In combination with the fact that OpenAI has yet to release the source code to GPT-3, this means that detection of GPT-3 generated text is made exceedingly difficult compared to if the source was available, even if one can use an equivalently large discriminative model for detection instead of GPT-3 itself.

\subsubsection*{The Issue with Simple Detection}

Concerns have been brought up in the literature that attempts to detect machine-generated text are less useful than they initially appear. Specifically, there is a glaring issue that concerns the use of machine-generated text detection approaches to prevent malicious use of generative LLMs: the assumption that use in general and malicious use are one and the same. One can easily conceive of several legitimate uses of machine-generated text via LLMs: as a writing aid tool, for automatic summarization of long texts, etc. All approaches that seek to simply detect use of LLMs do not distinguish between malicious and lawful use of these models \cite{schuster_limitations_2020}. Furthermore, most approaches that focus on simply detecting generated text can be bypassed by simply a) using more massive LLMs that provide more realistic text (e.g.\ GPT-3), but also b) using simple statement inversion and other semantically equivalent modifications to confuse the detector \cite{schuster_limitations_2020}. In certain experiments, models like GROVER have been used to demonstrate that although models can distinguish generated texts from human-written texts, they are still unable to separate \textit{truthful} content from \textit{misleading} content (for example, texts with strategically introduced negation to make them untruthful) \cite{schuster_limitations_2020}. This intuitively makes sense, as being able to distinguish these two elements requires knowledge about the world (i.e. knowledge-graph based approaches).

Alternative approaches have been somewhat effective at discriminating generated text from real text, albeit in somewhat limited contexts. A bidirectional LSTM in combination with a CNN layer has been used to model a notion of ``semantic coherence'', under the assumption that generated text will typically have errors that real text does not (e.g. incorrect word order). By perturbing real texts by swapping certain words randomly, one can simulate what a generated text could look like and create a synthetic dataset for detection \cite{bao_learning_2019}. However, these approaches have limited applicability to GPT-3 and other models of similar quality, as they are much more effective at emulating the structure of real text, and only very rarely make such errors.

\subsubsection*{Detection of Fake News Content}

Since simply detecting machine-generated text has the aforementioned problem of distinguishing legitimate from illegitimate use, a much more relevant question is how to detect fake news that has been generated from a LLM such as GPT-3. The following section seeks to answer this question by giving an overview of fake news detection approaches, and critically analysing which approaches would fail under generated fake news from an LLM model, and which would still function. The most obvious approach, the utilization of Knowledge Graphs to verify the actual veracity of a piece of news content, unfortunately is also the most infeasible, especially given early detection scenarios. Knowledge graphs are frequently manually curated, and require time to integrate new information (e.g. regarding novel events), and as such, are also simply inherently limited in scope \cite{zhou_fake_2020} \cite{liu_fned_2020}. The sheer volume of news generated each day means that KG-based approaches cannot possibly keep up \cite{liu_fned_2020}. Fake news has been demonstrated to be able to cause significant damage within extremely short timespans (a few hours), while KG-based approaches potentially require days \cite{liu_fned_2020}. Fact checking sites (Snopes, Politifact, etc.) are susceptible to the same problem: it takes time to identify the news and add it to the site, and in the meantime the damage may already be done \cite{liu_fned_2020}.

\vspace{5mm}

Approaches for detecting fake news can be divided into the following broad categories:

\begin{itemize}
    \item Content-based approaches
    \item Social-response-based approaches
    \item Hybrid approaches
    \item Graph-based approaches
    \item Multimodal approaches
\end{itemize}

\subsubsection*{Issues of Comparison and Dataset Standardization}

Unfortunately comparing approaches between papers is somewhat difficult. Unlike in the computer vision domain, there is no common benchmark dataset like ImageNet that allows for standardized comparisons between papers \cite{guo_future_2020}. Instead, papers typically craft their own datasets, frequently from the same or similar social media sources. Two common sources are Twitter and (Sina) Weibo, a Chinese social media platform. Other sources like Snopes, Politifact, and Buzzfeed are often used for labelling \cite{guo_future_2020}. Furthermore, there are two different goals to optimize for: the predominant goal of most literature in the domain, which is optimal detection ability irrespective of time eclipsed; and early detection, which has recently gained more attention as a more important goal, given the damage that fake news can do in a brief amount of time \cite{liu_fned_2020}.

\subsubsection*{Content-based Approaches}

Pure content-based approaches have been employed to some success. These approaches entail feeding the news article content itself to an algorithm to generate a binary label of fake/real news. It has been shown that fake news frequently has distinctive stylistic elements at various levels (semantic, lexical, discourse) \cite{zhou_fake_2020} \cite{kwon_rumor_2017}. Some of these elements include informality, lack of diversity of vocabulary, and use of emotionally charged words. In headlines specifically, the sentiment polarity and ``clickbait''-like format of a headline has shown to correlate with fake news \cite{zhou_fake_2020} \cite{shu_fake_2017}. Models based on a hand-crafted set of stylistic-element-based features have shown to be effective \cite{zhou_fake_2020}. These types of models do not rely on propagation information or other external features aside from the content itself, and thus are the naive best fit for early detection. There are certain issues with them at the same time: it is hard to prove generalizability, given that style can vary quite a bit depending on the area of focus of the fake news. These models are inherently limited in this way, and are likely unsuitable for machine-generated fake news detection. This is because LLMs do not have a notion of ``real'' and ``fake'' news: they simply generate text. This means that they do not expose ``tells'' that they are producing untruthful text, unlike humans, as the LLM itself does not know the difference. Furthermore, an LLM is likely to be very effective at mimicking the style of actual true news content, and thus would not necessarily exhibit the characteristics of fake text that these approaches use for discriminating fake text from real text \cite{schuster_limitations_2020}.

\subsubsection*{Social-response-based Approaches}

An alternative approach that is fairly common in the literature is to amass user responses to a certain event or news item, and feed these in aggregate to various ML algorithms. User responses are typically composed of short texts that convey the user's stance on the news item. RNNs are a common model to employ for such approaches, due to their ability to model changes over time. The user responses are typically fed to the RNN in sequence, with the hidden vector passed to a final fully connected layer for the binary prediction. Early work used TF-IDF values of a limited vocabulary \texttt{k} as input to the RNN \cite{ma_detecting_2016}, and was shown to be effective. Further research applied attention layers to enhance interpretability and allow the model to attend to specific phrases more effectively. Both approaches in essense mine the responses of users for terms like ``fake'', ``untrue'', etc., taking advantage of the ability for users to recognize fake content on their own \cite{ma_detecting_2016} \cite{chen_call_2017}. Furthermore, expressions with higher emotionality (e.g. ``unbelievable!'') in the responses of users were shown to correlate with fake news, likely owing to the fact that fake news is written in a way to purposefully evoke an emotional response. The presence of emotional terms in the responses of users was shown to correlate more closely with fake news than references to the news item content itself \cite{chen_call_2017}. Certain political terms as well seem to correlate with fake news (e.g. ``PC shit'' [sic], where ``PC'' is short for ``political correctness'') \cite{chen_call_2017}. It is likely that metadata relating to geographic location and IP addresses could also be important features in building a classifier to detect fake news in practice; not much academic research exists on using this metadata as features, as it is data that is not generally publicly available. It is very plausible that e.g. users in certain countries would be more involved in generating fake news content than others; it is generally well known that many misinformation campaigns arise from IP addresses associated with specific non-Western countries, be it individual hacker groups or even government-orchestrated campaigns.

\subsubsection*{Hybrid Approaches}

Several approaches seek to take the social response from users and augment it with certain user characteristics. CSI (Capture, Score, Integrate) \cite{ruchansky_csi_2017} represents the first attempt to unify the social response with characteristics of the users who are responding to a given news item. CSI is composed of three modules: the capture module, the score module, and the integrate module. The capture module is the same as other social-response-based approaches, and is simply an RNN that is fed the user responses to a given news item in a sequential fashion (encoding the text using doc2vec). The score module creates a global user vector per user via an incidence matrix representation of the news items they've interacted with, as well as a corresponding suspiciousness score. The suspiciousness score is the combination of two components: a vector representation of each user based on a set of user features passed through a single connected layer, and the result of applying single value decomposition (SVD) to the adjacency matrix representing the entire social network of users. The idea is that suspicious users likely interact more with other suspicious users. The score module then sums up the suspiciousness scores of the users who've interacted with a given news item to create a news-item-level score. This score is then concatenated by the integrate module to the capture module's output for the final prediction. The user vector can hypothetically be used for other analysis tasks, in combination with the global per-user suspiciousness score. The issue with this approach is that the score module is quite complex, having to analyse all of the news items every user has interacted with to generate the global suspiciousness score. Furthermore, it assumes that users have a steady ``suspiciousness'' over time, and loses the nuance that some users might sometimes share fake news, but not always.

Another approach seeking to instead integrate the content with the user responses is known as dEFEND \cite{shu_defend_2019}. This approach uses a \textit{hierarchical attention neural network} \cite{yang_hierarchical_2016} to examine the content of the news item at both a word and sentence level. A co-attention layer seeks to quantify the correlation between sentences in the news item content and specific user responses, to provide explainability through visualization of the strongest correlations between news content sentences and user comments. These most important sentence-comment pairs allows the model to isolate the exact sections of a news item that are fake (with the understanding that in fake news it is not the entire news item that is fake, but only certain parts of it interspersed with truth), and the user commentary that relates to those specific parts. This approach represents a large step in explainable fake news detection.

FNED \cite{liu_fned_2020} is a SoTA approach that builds off of the CSI approach to simplify it and focus on early detection specifically. It aims to fix the assumption of CSI that users can be boiled down to a single suspiciousness score that does not vary over time. Instead of generating such a score taking into account all interactions of all users in the dataset, FNED uses a convolutional layer to extract textual features from the response, then concatenates this with the result of passing the user features through an embedding block. This concatenated vector is called the ``status-sensitive crowd response feature'', a direct merger of the response text itself and the user information, and is fed to a CNN to perform the actual discriminative task. The CNN is augmented with a ``position-aware attention layer'', a regular CNN attention layer that is augmented with the position of the response within the response series, with the assumption that how early a response is matters in terms of detection. The approach is demonstrated to be extremely effective, especially in an early detection context.

\subsubsection*{Graph-based Approaches}

Other approaches seek to model the differences between fake and real news in terms of the propagation patterns through the network as news items are shared, and the relationships between various entities in the social network graph. The hypothesis is that fake news propagate substantially differently through the network than real news. One approach, known as TraceMiner \cite{wu_tracing_2018}, uses RNNs in combination with graph embedding algorithms to model the social graph in a way that can be fed to the RNN. The embedding algorithms used are SocDim, LINE, and DeepWalk (\cite{tang_relational_2009} \cite{LINE} \cite{Deep_Walk}). These embeddings produce user vectors that represent the structure of the network surrounding a given user. This approach consists of first learning the user embeddings via one of the aforementioned algorithms, then feeding the embeddings of the users who responded to a given news item to the RNN in sequence, similar to how the content of the responses are fed to RNNs in the social-response-based approaches. This approach has been demonstrated to be competitive with social-response-based approaches, which are more prone to manipulation by malicious spreaders \cite{wu_tracing_2018}. This approach also is quite effective in early detection scenarios, or scenarios where response information is quite limited (on most platforms, users have the ability to simply share the story without commentary).

Another successful approach is known as FANG \cite{nguyen_fang_2020}. FANG uses GraphSAGE (\cite{hamilton_inductive_2018}) to create user embeddings to feed into a hybrid BiDirectional-LSTM Attention-based model. RoBERTa is used to detect each user's stance to a particular news item they interacted with (stance detection) based on their response text. The embeddings generated by GraphSAGE take into account the users (characterised by their profile text, which has been shown to be a reliable indicator of a propensity to share fake news \cite{nguyen_fang_2020}), the news item content (a fusion of GloVe embeddings and TF-IDF scores), the character of the publishers (represented by TF-IDF scores of their About Us page and Homepages), and finally the relationships between all these entities, including which users follow which other users. Fed into the Bi-LSTM are the GraphSAGE representation of the user, their stance towards the given news item, and the time elapsed since the news item's publication. To create a final prediction, the GraphSAGE embedding for the news source and the Bi-LSTM output are concatenated together and fed to a fully-connected layer. The approach has been demonstrated to be highly effective compared to other graph-aware approaches. Furthermore, the Attention layer indicates that for fake news, the model attends to responses mostly within only 12h after publication, while for real news, the attention is spread out much more evenly over a two week period. This intuitively makes sense, as fake news is designed to cause immediate outrage, and thus propagates quite quickly, while real news circulates and remains relevant for longer spans of time.

Other approaches focus on injecting more information into the graph embeddings. The ideological slant of the publisher of a given news item on a left-right spectrum has been shown to improve the accuracy of graph-based fake news detection approaches \cite{shu_beyond_2018}, when modelled jointly with user interactions and a user adjacency matrix. Other approaches, such as GCAN (Graph-aware Co-Attention Network) \cite{lu_gcan_2020}, seek to improve interpretability of DL approaches for fake news detection. GCAN integrates several different feature extractors together into one network: an RNN (GRU) layer that encodes the source tweet (the news headline), the propagation pattern as an input to a convolutional layer, the propagation pattern as an input to a GRU layer, and finally, a Graph Convolutional Network (GCN) layer to generate an embedding for a given source tweet by modelling the neighborhoods of users who interacted with the tweet in a constructed graph. The propagation pattern inputs to the GRU and convolutional layers are simply a set of handcrafted user features (number of followers, length of description, etc.) fed in sequence of interaction with the original tweet to the layers. GCAN was demonstrated to improve the SoTA scores significantly, and does not rely on the actual replies of the users (who frequently repost without adding commentary), but rather simply the users themselves and the order of interaction. GCAN also integrates a ``dual co-attention'' mechanism to learn attention maps between both the graph-aware (GCN) representation of the interactions and the source tweet encoding, and the source tweet encoding and convolutional propagation encoding. This confirms the results of previous research (\cite{chen_call_2017} \cite{ma_detecting_2016}) that use of certain emotionally charged words in the source tweet correlates with fake news, as well as certain user characteristics like short account descriptions and recent account creation time. The latter implies malicious actors use bots and/or create fake accounts specifically for the promotion of fake news (and thus these accounts are young).

\subsubsection*{Multimodal Approaches: Incorporating Visual Information}

In recent years, multimodal approaches that aim to incorporate information from multiple modalities into the predictive model have gained more and more popularity. Most multimodal approaches concern the fusion of visual and textual information. It has been demonstrated that the use of images related to real events present far more variability in terms of content than those related to fake news \cite{jin_novel_2017}. The use of hand-crafted visual features to quantify the distinctiveness of images has been demonstrate to boost baseline accuracy by up to 7\% \cite{jin_novel_2017}. However, it is very time-consuming to craft features by hand, and so other, more recent approaches seek to incorporate visual information using existing feature extractor convolutional networks like VGG. Early research incorporating visual information via a VGG component in conjunction with word embeddings for actual post content and hand-crafted social (user characteristics) features demonstrated a marked improvement compared to approaches without the visual features \cite{jin_multimodal_2017}.

One markedly different approach is known as EANN (Event Adversarial Neural Networks) \cite{wang_eann_2018}. This approach utilizes a GAN-like training process to create a model which is hypothesized to be more generalizable compared to existing approaches. Based on a hypothesis that existing approaches rely on characteristics specific to the events present in the dataset they are trained on (are ineffective at generalizing), EANN is trained to extract event-agnostic features with which to perform classification. The architecture is similar to other approaches which incorporate VGG and textual features, however, the training is modelled as a minimax game between the feature extractor (textual + VGG) and an event discriminator. The event discriminator is trained to determine which event is being fed to the feature extractor from the concatenated visual and textual representation the feature extractor emits. In a similar fashion to GAN training, the feature extractor is forced to learn a representation which betrays as little event-specific information as possible, to fool the discriminator. This approach was demonstrated to exhibit a marked improvement in accuracy compared to the SoTA of the time. 

Another relatively different approach is known as MVAE (Multimodal Variational Autoencoder) \cite{khattar_mvae_2019}. MVAE uses a bimodal variational autoencoder to incorporate both text and image information into the predictive process. Under the hypothesis that existing approaches do not find correlations across modalities, MVAE was designed to incorporate a single shared representation over both textual and visual modalities for the reconstructive process. Bidirectional LSTMs are used to generate the textual component of the shared representation, while VGG-19 is used for the visual feature extractor. These representations are concatenated together to create the shared latent vector, which is then used to separately reconstruct both the original text and the VGG-19 extracted features. This approach was demonstrated to significantly outperform other multimodal approaches like att-RNN (\cite{jin_multimodal_2017}) and EANN (\cite{wang_eann_2018}), while not utilizing any propagation information at all, and operating only using the content of the news item itself (the orignal text and associated image(s)). This makes MVAE an especially good candidate for early detection.

\subsubsection*{Potential Research Direction \# $\ccc$ (Fake News Detection)}

There is no existing research covering multimodal detectors that incorporate explicit graph representations (not just sequences of user responses). It is conceivable that a model that incorporates VGG-based image features, content text, user response text, \textit{and} relationships between users (via DeepWalk or other similar graph embeddings) would be even more effective at detecting fake news than the approaches outlined here. Furthermore, most approaches that utilise profile features typically hand-craft them based on what the authors believe may correlate with users who propagate fake news. However, a potential new direction may be to attempt to create profile-based ``user embeddings'' (in contrast to FANG's GraphSAGE-based user embeddings) via passing user information (e.g. recent tweets) through a language model like Sentence-BERT. The intuition behind this would be that a user is likely best defined through their most recent tweets, given that opinions (and by extension the propensity for a given user to spread misinformation) change over time.

\newpage

\section{Topic and Sentiment Modelling for Social Media} \label{chp:topsent}

\subsection{Introduction}

This chapter presents an overview of topic and sentiment analysis approaches, as applied to social media posts (such as on Facebook or Twitter). We outline certain challenges relating to sentiment analysis as a whole that cause it to be a somewhat challenging problem, as well as challenges specific to social media platforms that make both topic and sentiment analysis more challenging than in the usual cases. An overview of classical topic modelling approaches, including  Latent Dirichlet Allocation (LDA), as well as newer, more modern approaches for topic modelling that incorporate deep learning is provided. Various sentiment analysis methods are also discussed, split into two major categories: unsupervised rule-based methods, involving the creation of hand-crafted rules for modelling sentiment, and the current SoTA in semi-supervised (transfer-learning based) approaches, which typically leverage massive pre-trained language models like BERT, RoBERTa, and various other Transformer-based architectures. These approaches involve unsupervised pre-training on massive amounts of unlabelled data (or, more typically, simply using an existing publicly available pre-trained model), followed by few-shot prediction after fine-tuning on a small amount of labelled examples. Finally, a brief overview of two new and difficult dimensions of sentiment analysis is given: multimodal sentiment analysis, which pertains to the analysis of the sentiment of multiple modalities at once (e.g. image, audio, and video data, especially relevant to social media posts, which frequently include images and video along with text), and target-based sentiment analysis, which involves detecting both the sentiment and the target of said sentiment in a piece of text. Methods for analyzing sentiment over time are briefly discussed.

\subsection{Introduction to Topic Modelling}

The goal of topic modelling is to discover a set of ``topics'' (abstract concepts/categories) that exist in a document corpus. Specifically, the goal of a topic model is to a) discover the existing topics in a corpus, typically modelled as latent variables, and b) to assign each document some specific mixture of topics, generally modelled as a proportion in relation to each of the discovered topics (X\% topic A, Y\% topic B, and so on and so forth). Similar to most clustering algorithms, where the user must pick the number of clusters before applying the algorithm, with most topic models the user must set a specific number of topics to be discovered.

\subsection{Overview of Classical Approaches to Topic Modelling}

\subsubsection*{LDA}

Latent Dirichlet Allocation (LDA) is a common generative approach to topic modelling that involves unsupervised discovery of latent topic variables based on a corpus of documents. Specifically, each topic is defined by a collection of words that represent it. Each word belonging to each topic has an associated probability (posterior probability), which represents the likelihood of that word being selected in the generative process given the topic's presence in a document. For the LDA model to be utilized, the posterior distribution first needs to be learned from the training corpus. The number of topics to be discovered is preset at the beginning of the process, similar to the selection of a number of clusters for an unsupervised clustering algorithm.

The generative process involves several steps:

\begin{enumerate}
    \item For each document generated, a topic distribution is drawn based on the Dirichlet distribution (parametrized by vector $\alpha$)
    \item For each word in the document to be generated (excluding stop words and common words), a topic is sampled from the topic distribution
    \item Finally, a word is sampled from the probability distribution of words belonging to the previously sampled topic
\end{enumerate}

The problem with LDA is that the true posterior distribution is intractable, and thus sampling methods are used (collapsed Gibbs sampling, Monte Carlo EM), but these are are computationally very expensive. This makes it difficult to perform interactive topic discovery, as each change to the modelling assumptions or even changes to the document corpus itself require a costly recalculation of the posterior distribution.

Another challenge with topic modelling is the difficulty in evaluating the generated topics. Most papers use a variety of different methods for evaluation: unfortunately, there is no ``standardized'' metric, nor a single standardized dataset for evaluating topic models. Human evaluation is effective but costly. One way that is relatively frequently used is normalized (pointwise) mutual information (NPMI). NPMI essentially models the ``coherence'' of topics, typically averaged across all topic words. A higher average NPMI means that the words within the topic category co-occur at a higher rate, and thus generally indicates high topic coherence. NPMI has been demonstrated to be correlated with human evaluation of topic quality \cite{aletras_evaluating_2013}.

\subsection{Neural Topic Modelling}

\subsubsection*{Variational Topic Modelling}

Neural topic models (NTMs) aim to perform unsupervised topic discovery from a corpus of documents using neural networks. One approach that aims to combine LDA and the autoencoding variational Bayes (AEVB) \cite{kingma_auto-encoding_2014} architecture is known as Autoencoded Variational Inference For Topic Model (AVITM) \cite{srivastava_autoencoding_2017}. AEVB avoids the expensive computation of traditional methods for learning the posterior distribution by training a neural encoder network to learn the distribution directly. AVITM tackles two issues regarding the use of AEVB with a Dirichlet prior: a) the common problem in the AEVB of component collapse, where the network gets stuck in a bad local optimum, and b) the issue of reparameterizing the LDA prior (which is necessary to use the AEVB approach), which is challenging because the Dirichlet family of distributions are not a location-scale family. AVITM solves the issue of component collapse by using a combination of the Adam optimizer with a high learning rate and high momentum, batch normalization, and dropout units to avoid getting stuck in local optima. The reparameterization of the Dirichlet prior is solved by approximating it via a logistic-normal distribution. The result is a black box inference method that learns the prior distributions (via the variational parameters) necessary for LDA via neural network (the ``inference network''), accepting as input the documents of the corpus (the ``observed data'') for training. The ``autoencoding'' nature of AVITM arises from the fact that the inference network is composed of two parts: an encoder and a decoder, where the encoder takes a document and produces a continuous latent representation for it, and the decoder takes the latent representation and attempts to reconstruct the original words from the BoW representation of the document (essentially performing the LDA generative process). The proposed approach is far faster than traditional sampling approaches, and is feasible to run on a one million document dataset in under 80 minutes on one GPU. Furthermore, AVITM provides much faster inference on new data compared to other approaches (like standard mean field), as it is simply a matter of passing the new data point through the neural network. Furthermore, in the same paper as AVITM an approach called ProdLDA is proposed, which produces better and more coherent topics than the standard LDA approach, by replacing the word-level mixture model of LDA with a weighted product of experts model.

A similar approach to AVITM is known a the Neural Variational Document Model (NVDM) \cite{miao_neural_2016}. The primary differences with AVITM is that instead of reparameterizing the Dirichlet prior, a latent Gaussian distribution for the topics is assumed instead, which allows for more straightforward reparameterization, but carries a certain bias. As well, the high momentum Adam training is not used to avoid component collapse.

\subsubsection*{LDA2Vec}

\begin{figure}[ht]
\centering
\includegraphics[angle=0,width=6cm]{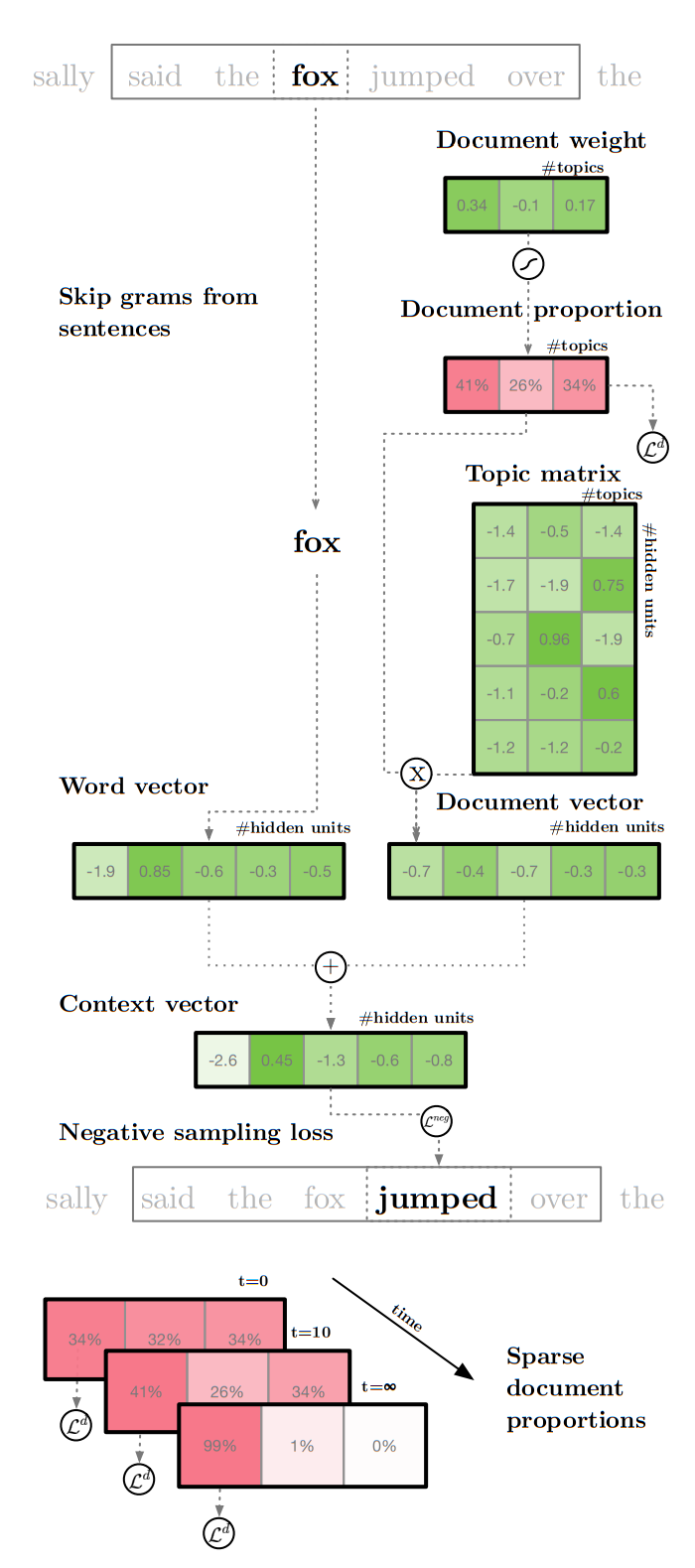}
\caption{The pre-training process for LDA2Vec. The addition of the document vector and its contribution to the pre-training task (the negative sampling loss) is clear. \cite{moody_mixing_2016}}
\label{fig:lda2vec}
\end{figure}

LDA2Vec \citep{moody_mixing_2016} represents a different approach to NTMs, seeking to augment the Word2Vec model by adding a learned document representation to Word2Vec's Skipgram Negative-Sampling (SGNS) objective, in addition to the context-independent word vectors that are learned. These document vectors are the product of a document weight vector, which represents the proportion of each topic present in a given document, and the topic matrix, which represents topics as points in the word embedding space (though they do not necessarily correspond to actual words). The fact that both topics and words exist within the same embedding space allows topics to be easily visualized and understood via the words in their neighborhood in the embedding space. The existing SGNS objective is not modified: the pre-training task is simply predicting for a given pair of words whether one follows the other. The idea of the modified SGNS objective of LDA2Vec is that document information can help in the pair classification task, and therefore result in more robust word vectors. In regular Word2Vec, summing the vector for ``Germany'' and the vector for ``airline'' may result in a similar vector for ``Lufthansa''. In LDA2Vec, a document about airlines would have a similar document vector to the word vector for ``airline''. As the negative sampling loss is calculated in relation to a ``context vector'', which is the sum of the word vector and the document vector for the current document, the classification of the word pair is rendered more effective, with this additional document-level information (see Figure \ref{fig:lda2vec} for a visual representation). These three representations (word, document, topic) are jointly discovered during the pre-training process via the modified SGNS objective.

\subsubsection*{Top2Vec}

Top2Vec \citep{angelov2020top2vec} is a similar approach to LDA2Vec, that however differs in a few key ways. Both LDA2Vec and Top2Vec seek to model documents as vectors within a shared word/ document embedding space. LDA2Vec jointly generates interpretable document topic-mixture representations (a vector with the proportions of each topic that compose the document) via the modified pre-training objective in addition to the final document vector in the word embedding space. In contrast, Top2Vec opts for the comparatively simpler and more straightforward approach of performing clustering directly on the word/document embedding space via the HDBSCAN algorithm after applying dimensionality reduction. The word/document embeddings can be generated by any algorithm that produces such joint embeddings, as long as both documents and words are jointly embedded in a shared semantic embedding space. Some examples are Doc2Vec, Universal Sentence Encoder, and BERT Sentence Transformer, the implication being that the quality of these shared embeddings will determine the performance of the approach to a large extent. The centroid of each cluster in the shared embedding space as produced by HDBSCAN is considered to be the ``topic vector'', and the 5 nearest word vectors to the topic vectors are taken as the topic descriptors. The distance from each topic centroid to each document vector in the embedding space could hypothetically be used to produce a similar topic-mixture representation as LDA2Vec, however this is not a key part of the algorithm in the same way as it is in LDA2Vec.

\subsubsection*{Use of Pre-trained Embeddings for Neural Topic Modelling}

Recently the use of pre-trained embeddings leading to SoTA results for many NLP tasks has prompted researchers to begin to look for ways to incorporate them into unsupervised topic modelling. One approach to doing so is to extend the AVITM approach by augmenting the traditional BoW representation with contextualized Sentence-BERT embeddings \cite{bianchi_pre-training_2020}. The sentence embeddings of the sentences in the document are concatenated to the original BoW representation of the document before passing the result of the concatenation to the autoencoder, allowing the autoencoder to use the BERT embeddings to aid in the reconstruction of the BoW representation of the document. The BERT embeddings are first passed through a single hidden layer prior to concatenation. Higher topic coherence is demonstrated compared to the original ProdLDA, as well as both NVDM and traditional LDA, indicating that the injection of external information is beneficial to the topic discovery process.

Another approach to topic modelling using pre-trained embeddings eschews the use of probabilistic models entirely, instead clustering vectors in the embedding space directly from pre-trained embeddings like BERT, GPT-2 and RoBERTa, with the intuition that proximity in the embedding space between two word embedding vectors suggests a common theme \citep{thompson_topic_2020}. Specifically, k-means can be used to produce word clusters (topics) that resemble the prior distribution produced by LDA. The use of contextualized BERT embeddings allows the clustering approach to better account for polysemy (different meanings of the same word) compared to the topics formed by traditional LDA, which does not have access to contextual information (as it uses a BoW representation). The contextualized embeddings also capture more varied parts-of-speech to represent each topic, unlike LDA which generally prefers nouns. By using contextualized embedding clustering, analyses over partitions of the corpus can be performed, for example estimating the prevalence of certain topics over time, in the case of a corpus with temporal information associated with documents. Furthermore, it is demonstrated that the clustering in the embedding space is meaningful to produce topics even when PCA dimensionality reduction is applied to reduce the 768-dimension BERT embeddings down to 100 dimensions.

\begin{figure}[ht]
\centering
\includegraphics[angle=0,width=8cm]{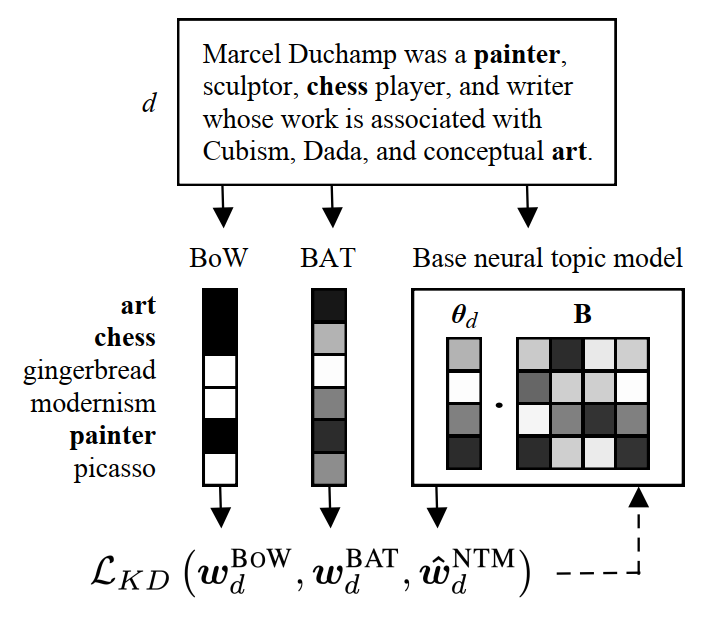}
\caption{An illustration of the BoW representation of a document alongside the BAT representation. Weight is given to related terms that are not actually present in the document, improving the topic model. \cite{hoyle_improving_2020}}
\label{fig:bert-bat}
\end{figure}

A somewhat different approach is improving existing neural topic models with pre-trained transformer networks via the process of knowledge distillation \cite{hoyle_improving_2020}. This approach is named ``BERT-based Autoencoder as Teacher'' (BAT), and involves fine-tuning a pre-trained BERT model with a document reconstruction objective, to act as an autoencoder. The BAT representation of a document, unlike the BoW representation, incorporates related but unseen terms into the representation distribution (vector), informed by BERT's pre-trained knowledge (see Figure \ref{fig:bert-bat}). The BAT representation is incorporated as a KD term in the loss function for the NTM, which guides it to mimic the BAT representation. This allows the NTM to incorporate the knowledge contained within the the pre-trained BERT embeddings while maintaining the interpretability inherent to the NTM approach.

\subsubsection*{Neural Topic Modelling for Social Media}

Social media creates additional challenges compared to corpora with longer documents when it comes to the application of traditional LDA approaches. LDA produces topics based on the co-occurence of topic words within documents. However, short documents (social media posts) create much sparser co-occurence statistics, making topic discovery more challenging.

One approach that uses BERT pre-trained embeddings for topic classification (not discovery) for social media posts specifically is known as TopicBERT \cite{chaudhary_topicbert_2020}. TopicBERT leverages the NVDM architecture for document classification, rather than discovery. TopicBERT concatenates the NVDM latent representation with the BERT word embeddings to perform more effective document classification, and achieves higher scores in several benchmark document classification datasets (Reuters8, 20 Newsgroups, and IMDB), compared to a set of baseline architectures (plain CNN, BERT, DistilBERT). This demonstrates that the NVDM latent representation can provide additional information to boost document classification accuracy in a social media context.

Another approach that more closely aligns with traditional LDA is that of the Latent Concept Topic Model (LCTM) \cite{hu_latent_2016}. Unlike LDA, which models each topic as a distribution over words in the document corpus, the LCTM instead seeks to model each topic as a distribution of latent concepts. These concepts are modelled as Gaussian distributions in the pre-trained word embedding space (in the LCTM case, GloVe vectors). The use of latent concepts (modelled as ``concept vectors'' in the word embedding space) instead of using words directly helps to ameliorate the sparsity problem for short texts, as the number of distinct concepts in a corpus is far fewer than the number of distinct words. The co-occurence statistics of latent concepts are therefore much less sparse. A collapsed Gibbs sampler approach is used for inference. The LCTM is also able to take advantage of OOV terms during inference on new data, by mapping them to the existing latent concepts via the embedding space. The LCTM has been demonstrated to be far more effective than traditional approaches at modelling topics in short texts (e.g. on the 20 Newsgroups dataset).

A similar approach to LCTM is known as GPU-DMM \cite{li_topic_2016}. GPU-DMM uses a Dirichlet Multinomial Mixture (DMM) architecture augmented with pre-trained Word2Vec embeddings to group words with similar meanings in a similar way to LCTM. DMM is a similar approach to LDA, but instead assumes that a document is generated from a single topic, rather than a mixture of topics. For short texts, this is generally a fair assumption to make. The second component of GPU-DMM is the Generalized Polya Urn Model (GPU). The GPU allows the DMM model to incorporate semantically related words into the sampling process (which is similar to collapsed Gibbs sampling), where these semantically related words are based on proximity in the word embedding space. This allows it to achieve better results on short texts compared to traditional LDA, by in effect increasing the size of the texts being learned from.

The Context Reinforced Neural Topic Model (CRNTM) approach modifies the NVDM architecture to be more effective for short texts \citep{feng_context_2020}. It does this by modelling topics as Gaussian mixture distributions within the pre-trained word embedding space, to alleviate the issue of feature sparsity with short texts. In a similar way as LCTM, CRNTM seeks to enrich the contextual information in short texts via nearby words in the embedding space. Unlike LCTM, this approach is used in combination with a more modern VAE-NTM approach, rather than a traditional sampling process.

\subsubsection*{Potential Research Direction \# $\ccc$ (Extending NTMs)}

Generally speaking not as much research has been done on topic modelling for short texts, ergo this seems like the area with the most potential for novel research. Bidirectional transformer representations (e.g. BERT) have been used to perform topic modelling via clustering the learned document representations (e.g. \cite{hoyle_improving_2020}). However, the potential for large autoregressive models to perform one-shot or zero-shot topic modelling has been relatively unexplored. One novel line of research could be using GPT-3 and in-prompt learning to perform topic modelling on a short text corpus without gradient descent. One limitation would be that a representative sample of the corpus would have to fit into the prompt-length of the model, however there is much potential for the use of proper prompt tuning (a newly emergent field relating to optimizing in-prompt learning for large autoregressive models by carefully selecting prompts) to aid in effective topic modeling.

\subsection{Sentiment Analysis}

Sentiment analysis is not a single problem with well defined boundaries, but rather a family of problems relating to the detection of sentiment (emotion/attitude) in text. This family of problems can be roughly split into three categories:

\begin{enumerate}
    \item Document-based, which relates to the detection of the prevailing sentiment in an entire document (a series of sentences),
    \item Sentence-based, which relates to the detection of sentiment in a single sentence,
    \item and Aspect-based: \begin{enumerate}
        \item Simple aspect-based, where the goal is to detect sentiment relating to a certain ``aspect'' of the topic of the text (e.g. in a product review context, sentiment regarding price, quality, etc.)
        \item Target aspect-based, where the goal is to detect both the aspect, and the target of the sentiment as well
    \end{enumerate}
\end{enumerate}

Of these three cases, the most challenging and least-developed is aspect-based sentiment detection, which also involves the most moving parts, as both sentiment has to be detected, and potential targets of the sentiment have to be determined. These two subproblems are known as Aspect Extraction (AE) and Aspect Sentiment Classification (ASC). AE is a problem of sequence labelling, while ASC is a regular classification problem. These fundamentally different subproblems make it difficult to create a single uniform architeture that can perform both tasks. 

Sentiment analysis can furthermore be split into supervised and unsupervised cases. The supervised case can be modelled in several different ways. One of the most simple ways to model the problem is to use a simple polarity axis, to predict how positive or negative the sentiment is. More complex modelling strategies include trying to predict a specific emotion in a set of standard emotions (anger, sadness, disgust, etc.). The former can be set up as a standard binary classification problem, the latter as a multi-class classification problem where each emotion is a distinct class. When trying to model complex emotions as a multi-class problem, the Plutchik wheel of emotions is typically used (for more detail see Chapter \ref{chp:ontologies}).

\subsubsection*{Sentiment Analysis and Stance Detection Standardized Datasets}

In terms of standardized datasets for sentiment analysis, one common source is the International Workshop on Semantic Evaluation (commonly known as SemEval) \footnote{\url{https://semeval.github.io/}}. The datasets of SemEval cover a wide range of NLP problems, including sentiment analysis, as well as stance detection. Stance detection refers to a related problem to sentiment detection, which instead has as a goal to detect the user's stance with regards to a certain issue/topic (in favor, against, neutral). Stance detection is most similar to target-based sentiment detection in this way, in that a target of the emotion/stance is needed. SemEval 2016 specifically offered a dataset for stance detection in a set of 2900 tweets, where the goal is to detect the stance of a particular tweet with regards to 5 distinct topics (``Atheism'', ``Climate Change is a Real Concern'', ``Feminist Movement'', ``Hillary Clinton'', and ``Legalization of Abortion''), and 3 stances (``in favour'', ``against'', and ``none''). This dataset, despite having been used in several papers, is currently out of date, and the size of the dataset is not sufficient for more complex newer ML systems. A better, more recent option is the stance detection dataset of the Fake News Challenge \footnote{\url{http://fakenewschallenge.org}}, which consists of 49k rows. The dataset consists of headlines and body text from online articles. The goal is, for a given headline and body text (not necessarily from the same article), to classify the pair into one of 4 classes: ``agrees'' (the body text agrees with the headline topic), ``disagrees'' (the body text disagrees with the headline topic), ``discusses'' (the body text discusses but does not express opinion on the headline topic), and ``unrelated'' (the two are about unrelated topics).

\subsubsection*{Traditional Supervised Sentiment Analysis}

The supervised cases (excluding aspect-based detection) are relatively straightforward. BERT, RoBERTa, and other similar contextualized embeddings that have been pre-trained on massive amounts of unlabelled data provide a strong foundation for any supervised sentiment detection approach, through the use of transfer learning. Pre-trained weights for these architectures are available online. By using these pre-trained weights, these architectures can achieve SoTA accuracies via fine-tuning on fairly small labelled datasets, by leveraging their acquired knowledge through the pre-training process. At the time of writing, the current SoTA pre-trained embedding architecture is known as SMART-RoBERTa \cite{jiang_smart_2020}, and consists of a pre-trained RoBERTa model trained via a new learning framework that provides better generalizability to new data when fine tuning on small labelled datasets. A library known as HuggingFace \footnote{\url{https://huggingface.co/}} provides an easy-to-use high-level API for fine-tuning large Transformer-based NLP models for downstream tasks.

GPT-3 has been used for few-shot sentiment classification, leveraging the in-context training abilities of the model (mentioned in Section \ref{gpt-intro}). It has been demonstrated that the choice of in-context examples is a crucial factor in how effective GPT-3 will be for a given few-shot learning task, with highly variable performance depending on this choice \citep{liu_what_2021} \cite{zhao_calibrate_2021}. Specifically, it has been demonstrated that the more similar the in-context examples given are to the testing set used, the higher the effectiveness of the model, which intuitively makes sense. The KATE system \citep{liu_what_2021} uses a KNN-based sampling method for in-context examples, based on sentence embeddings produced by other models (e.g. RoBERTa), and demonstrates a consistent increase in accuracy compared to randomly selected in-context examples from the training set (which is the approach used for constructing in-context examples in the original OpenAI GPT-3 paper). KATE was evaluated on a sentiment analysis task (SST-2 \cite{socher_recursive_2013}) and demonstrated competitive accuracy, despite being trained in a few-shot context. Models like SMART-RoBERTa (current SoTA at the time of writing) achieve higher accuracy, but are fine-tuned with gradient updates and far larger training sets than can fit in GPT-3's model prompt for in-context (few-shot) training.

A popular example of a rule-based approach for entirely unsupervised sentiment analysis of text is known as VADER \cite{hutto_vader_2014}. VADER consists of hand-tuned word-sentiment correspondences (including sentiment intensity), in combination with rules that model how the presence of certain syntactical features modify the intensities (features like negation, degree modifiers, etc.). VADER is specifically tuned to the type of language used in social media contexts, and is aware of things like capitalization and punctuation, and how they modify sentiment intensity. At the time of publication of the VADER paper, VADER outperformed all ML sentiment analysis approaches, as well as human raters in certain cases. Most other fully unsupervised approaches use a variation of this lexicon-based strategy augmented with rules. One set of datasets that is often used to underpin such approaches is the NRC emotion and sentiment lexicons \cite{Mohammad13}.

\subsubsection*{Multimodal Sentiment Analysis}

Multimodal sentiment analysis refers to the use of multiple modalities for input for a sentiment (typically) classification model. There are two different ways that multiple modalities can be used: 

\begin{enumerate}
    \item Classifying sentiment based on not only the textual content of the post on social media, but also any images that have been uploaded with it, typically relating to the content which the post discusses
    \item Classifying sentiment based on multiple directly correlated modalities, typically speech utterances: given a source video of someone speaking, the audio of their voice, and a transcript of what they're saying, classifying their sentiment.
\end{enumerate}

Standard datasets for multimodal sentiment analysis primarily use the second problem structure, rather than the first. Standard datasets used include the Multimodal Opinion Utterances Dataset (MOUD) \cite{perez-rosas_utterance-level_2013} and the Multimodal Corpus of Sentiment Intensity (MOSI) \cite{zadeh_mosi_2016}. The first problem in the list above can be considered an area of open research, although it is also potentially a simpler problem, as correlations across modalities likely play a much smaller role than with utterance datasets. Nevertheless, it would be interesting to adapt one of the below models for the first task.

One approach to the second problem is known as MISA \cite{hazarika_misa_2020}. MISA uses separate feature extractors for each modality (BERT for text, bidirectional LSTM for image and speech) and projects each set of features per modality onto a distinct modality-specific subspace, as well as a single modality-invariant subspace. MISA is designed for synchronized multimodal classification: classifying video, speech, and text that all relate to the same context (a person speaking, where the video is of them speaking, the speech is the audio of them speaking, and the text is a subtitle). A Transformer model (with Multi-head attention) is used to fuse the different modalities together in the final stage to make a prediction.

Another new approach for the same speech-utterance task is known as TransModality \cite{wang_transmodality_2020}. TransModality extends the Transformer architecture to include ``modality-fusion cells'', which jointly learns the correlations across modalities by ``translating'' from one modality to the other, leveraging the Transformer's original design for textual translation. The TransModality model achieves SoTA results on the two standard datasets mentioned earlier.

\subsubsection*{Potential Research Direction \# $\ccc$ (Textual Sentiment Analysis over Time)}

There is very little existing research examining the problem of analysing sentiment over time in a textual corpus. One can envision two different approaches: one approach would be to leverage the transfer learning strategies outlined earlier, using a pre-trained model like RoBERTa to segment the dataset into set time periods, and then performing few-shot classification on each period to analyse sentiment in a corpus or dataset over time. A different approach might be to use time series based algorithms to attempt to capture the sentiment over time in a single model. The aforementioned speech utterance task does include a temporal component, as the dataset is composed of videos with per-frame labels of sentiment, so the models mentioned in the previous section can also be thought of as doing sentiment analysis over time, for the duration of the video. Analysis of entire textual corpora over time is a comparatively less explored research area. 

\subsubsection*{Aspect-based Sentiment Analysis}

Targeted aspect-based sentiment analysis (TABSA) is an especially challenging subproblem of sentiment analysis, composed of the separate AE and ASC subproblems mentioned earlier. One approach to ASC involves the use of an auxiliary sentence to reframe the problem as a sentence-pair classification task (similar to question answering) \cite{sun_utilizing_2019}. When compared to directly fine-tuning BERT on the task, the construction of an auxiliary sentence presents a substantial improvement to accuracy on the standard SentiHood dataset. Directly using BERT on the SentiHood task constitutes fine-tuning a separate BERT model for each aspect. The auxiliary sentence that is constructed takes the form of a question or statement relating one of the aspects and targets together (e.g. ``what do you think of the safety of location x''). It is hypothesized that the increase in accuracy compared to applying BERT directly to the task is due to the pre-training strategy of BERT, specifically the next sentence prediction component, that give BERT an edge when it comes to sentence-pair-based tasks.

A different approach to performing TABSA, instead of modelling the problem as a sentence-pair classification task, is to use conditional random fields (CRFs) in conjunction with the direct application of BERT without fine-tuning via layer aggregation \citep{karimi_improving_2021}. The CRFs inject aspect information (performing sequence labelling) into the training process which aids the network in performing the sentiment analysis. The last four layers of the BERT network are used for the aggregation, under the assumption that sentiment classification is a fairly high-level task, and thus the information required would be contained in these later layers. This is consistent with typical use in existing literature using BERT as simple embeddings for downstream tasks, where the output of the last 4 layers of BERT is used as input to regular linear layers trained on top of them, without gradient updates (i.e. fine-tuning) to the BERT layers themselves.

The success of BERT with regards to ABSA has been analyzed through the visualization of the fine-tuned BERT model self-attention heads, to see what features are being used in this classification task, and to interpret the latent space generally \cite{xu_understanding_2020}. Through this analysis, it has been shown that there is no single dimension or small set of dimensions responsible for the success of BERT when it comes to ABSA tasks. It has been demonstrated that the majority of the dimensions of BERT when fine-tuning on ABSA relate to the semantics of the aspect itself, rather than the opinions learned from the fine-tuning dataset. Nevertheless, a deeper understanding of why BERT is so effective at ABSA (and NLP tasks in general) remains somewhat elusive.

\subsubsection*{ASBA in a Unified Framework}

Most existing research in ABSA is focused on the classification subtask, with the assumption that targets have already been extracted. Despite this, there is some research that aims to focus on both subtasks simultaneously. GRACE (GRadient hArmonized and CascadEd labeling model) \cite{luo2020grace} is an end-to-end framework for performing both AE and ASC. It does so through a ``aspect term-polarity co-extraction'' framework, essentially performing both the AE and ASC task simultaneously. That is to say, provide two labels per token in a given sentence: one indicating if it is an aspect or not, and the second to provide the sentiment if it is. It seeks to solve the issues of a) class imbalance in the AE training set, in other words that the majority of the words in any given text are not aspects, and b) interplay between aspect terms in a given sentence, e.g. by coordinating conjunctions (``nice \texttt{a} and \texttt{b}'' should label both \texttt{a} and \texttt{b} as positive). The model uses a cascaded labelling approach, where the aspect labels are first generated from the pre-trained BERT embeddings, then are fed into a Transformer-Decoder block as the key and value parameters to generate the sentiment labels. The use of the Transformer-Decoder block (built on Multi-Head Attention) mitigates the issue of sentiment not propagating properly through coordinating conjunctions. The issue of class imbalance in the AE task is mitigated by using a gradient harmonized loss, borrowed from the problem space of object detection.

Another approach that aims to perform both AE and ASC in a unified framework is known as SpanABSA \cite{hu_open-domain_2019}. Most existing solutions to joint AE/ASC use a sequence tagging approach, where each word is given a term/not term label, as well as a positive/negative sentiment label (as mentioned previously). SpanABSA uses a different representation instead, where the target span is represented directly as a span (a start and end point in the sequence), rather than as a stream of tokens. As with GRACE, BERT is used as the DNN underpinning the framework. For the AE component, BERT is used to predict the start and ending positions (the span) directly, rather than predicting a label for each token in the input (as with most other sequence labelling approaches to AE). A heuristic approach is used to narrow down the suggested target spans based on several criteria, e.g. removing overlapping spans or softmax scores for the start and end positions below a certain threshold. These spans are then fed to the polarity classifier module (also BERT) to create the sentiment prediction. SpanABSA outperforms existing unified AE/ASC training frameworks, especially in the case of targets being more than 2 words long.

\subsubsection*{Potential Research Direction \# $\ccc$ (Aspect-based Multimodal Sentiment Analysis)}

The most active areas of SA with regards to new research are that of aspect-based sentiment analysis, and multimodal sentiment analysis. A natural path for future research could be the union of these two areas, that is to say: given a video of someone speaking, classify not only the sentiment displayed, but also the target of the sentiment. This task promises to be quite challenging, as it would require leveraging both span-based AE/ASC models (e.g. SpanABSA \cite{hu_open-domain_2019}), and modality fusion techniques from existing multimodal models. Despite being a natural extension of existing research areas, this problem is almost entirely unexplored as of the time of writing, perhaps due to its obviously challenging nature.

\newpage

\section{Mining and Modelling Complex Networks}\label{sec:networks}

Mining complex networks represented as graphs is now a well-established subfield within data science. Indeed, analyzing social networks by investigating metadata as well as the content produced by the users can be powerful but it does not capture the dynamics and various types of interactions between users. Such additional dimension can be naturally modelled as graphs in which nodes (associated with user accounts in the case of social networks) are connected to each other by edges (relations based on follower requests, similar user's behaviour, age, geographic location, etc.). Such networks are often large-scale, decentralized, and evolve dynamically over time.


Modelling complex networks using random graphs is a very active and popular area of research in pure mathematics, starting with a pioneering work of Paul Erd\H{o}s and Alfr\'{e}d R\'{e}nyi~\cite{Erdos1959randomgraphs} from 1959. The initial interest was mostly concentrated on investigating properties of the binomial random graph $\mathcal{G}(n,p)$ and random $d$-regular graphs but now the family of known and studied models include more realistic models of complex networks such as the Chung-Lu model~\cite{ChungLu2006book} or various geometric graphs such as the spatial preferential attachment model~\cite{aiello2008spatial} or the hyperbolic geometric graph~\cite{Krioukov2010}.  

Currently, we experience a rapid growth of research done in the intersection of mining and modelling of social networks. There are two main reasons to include random graph models in mining complex networks:
\begin{itemize}
\item \textbf{synthetic models}: Many important algorithms (such as community detection algorithms) are unsupervised in nature. Moreover, despite the fact that the research community gets better with exchanging datasets (see, for example, SNAP---Stanford Large Network Dataset Collection~\cite{snapnets}), there are still very few publicly available networks with known underlying structure, the so-called ground truth. Hence, in order to test, benchmark, and properly tune unsupervised algorithms, one may use random graphs to produce synthetic ``playground'': graphs with known ground truth (such as the community structure in the context of community detection algorithms). 
\item \textbf{null-models}: Null-model is a random object that matches one specific property $\mathcal{P}$ of a given object but is otherwise taken, unbiasedly, at random from the family of objects that have property $\mathcal{P}$. As a result, the null-models can be used to test whether a given object exhibits some ``surprising'' property that is not expected on the basis of chance alone or as an implication of the fact that the object has property $\mathcal{P}$. 
\end{itemize}
It is expected that both applications of random graphs will continue to gain their importance in the context of mining complex networks. There is a need for synthetic models of more general structures such as hypergraph as well as models that are dynamic (that is, they produce a sequence of graphs with both edge and node additions/deletions). Null-models are successfully used in designing clustering algorithms but they are expected to also play an important role in other aspects of mining networks, especially, as a predictive tool, when dynamics is captured in the form of a sequence of graphs or times when edges/nodes were added/deleted from the network. 

\bigskip

In the following subsections, we identify a few areas in which more research seems to be needed. 
The outcome of such research projects may potentially be of interest to the industrial world as well as academia at large. 
We refer the reader to the recent book~\citep{Kaminski2021book} for more standard applications of mining complex networks.

In each subsection, we briefly summarize current ``state of the art'' tools and approaches but also discuss various possible directions for future applied research. These parts will be clearly marked.

\subsection{Node Embeddings} \label{sec:embeddings}

An embedding is a function from the set of nodes $V(G)$ of some graph $G$ to $\R^k$, where $k$ is typically much smaller than $n$. In other words, the embedding represents each node as a low-dimensional feature vector. The goal of this function is not only to decrease the dimension but to also preserve pairwise proximity between nodes as best as possible. It is clear that embedding-based representations emerged and quickly increased attention over the last decade. As reported in~\citep{chen2021symbols}, the ratio between the number of papers published in top 3 conferences (ACL, WWW, KDD) closely related to Computational Social Science (CSS) applying symbol-based representations and the number of papers using embeddings decreased from 10 in 2011 to 1/5 in 2020.

A~closely related way to achieve this is to conduct relational learning by propositionalization. In this approach, relational information is first captured and stored as propositions, according to some predefined declarative bias. Propositional learning algorithms can then be applied to learn using these extracted features. Such algorithms, often probabilistic, are well studied and available to use. Finally, one may enrich the embedding algorithms by including some domain-specific relations constructed using Inductive Logic Programming (ILP). It seems that incorporating symbolic domain knowledge might improve the quality of algorithms and ILP can play an important role in providing high-level relationships that are not easily discovered by other means~\cite{dash2021incorporating}. 

\begin{figure}[ht]
\centering
\includegraphics[angle=0,width=6.5cm]{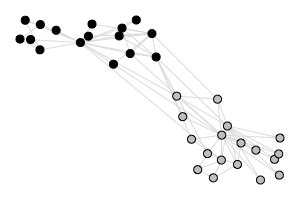}
\includegraphics[angle=0,width=6.5cm]{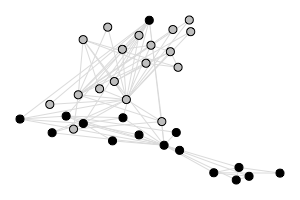}
\caption{Two-dimensional projections of two sample embeddings of the Zachary karate club graph.}
\label{fig:emb_zac}
\end{figure}

We present a toy example in Figure~\ref{fig:emb_zac}: an embedding of the Zachary's karate club graph, a popular example of a network with community structure, first used in~\cite{Girvan7821}. The graph represents social interactions between 34 members of a karate club; nodes of this graph are partitioned into two group as a result of a conflict between the club’s president and the instructor.

There are many ways the proximity can be measured: first, second, and in general $k$th-order proximities, Katz Index, Personalized PageRank, Common Neighbours, Adamic Adar, SimRank (see, for example,~\citep{Kaminski2021book} for the corresponding definitions). There are various known applications of node embeddings and many potential new directions are currently being explored, some of them will be mentioned later in this document. 
\begin{itemize}
\item Node classification is an example of a semi-supervised learning algorithm where labels are only available for a small fraction of nodes and the goal is to label the remaining set of nodes based on this small initial seed set. Since embedding algorithms can be viewed as the process of extracting features of the nodes from the structure of the graph, one may reduce the problem to a classical machine learning predictive modelling classification problem for the set of vectors.
\item Community detection can be reduced to (or, at least, supported by) clustering points in $\R^k$ which is a much easier task and is a well-studied area of research with many scalable algorithms, such as $k$-means or DB-scan, that are easily available for use. Some initial results (embedding + vanilla $k$-means or Gaussian mixture models) show a potential of this approach~\cite{tandon2021community}.
\item Link prediction algorithms (including predicting missing links or links that are likely to be formed in the future) can also be successfully built using node embeddings. Indeed, once nodes are embedded in $k$-dimensional space, one may use the distance between the corresponding vectors, combined with additional information, to build the corresponding null-model and to make the prediction. 
\end{itemize}

There are over 100 algorithms proposed in the literature for node embeddings. The techniques and possible approaches to construct the desired embedding can be broadly divided into the following families. For more details, the reader is directed to~\citep{hamilton2017representation}, \cite{zhang2018network}, and \citep{Kaminski2021book}.
\begin{itemize}
\item Linear algebra algorithms include Local Linear Embedding (LLE)~\cite{roweis2000nonlinear}, Laplacian Eigenmaps (LEM)~\cite{belkin2001laplacian}, High Order Proximity preserving Embedding (HOPE)~\cite{HOPE}, which is an interesting instance of this approach aimed at embedding nodes in directed graphs.
\item Random walk based node embedding methods are derived from the Word2Vec~\cite{Word2Vec} algorithm for word embedding commonly used in Natural Language Processing (NLP). Two representative algorithms from this family are Node2Vec~\cite{Node2Vec} and Deep Walk~\cite{Deep_Walk}. The common and general idea is as follows. The ``words'' are simply the nodes of a graph, and one generates ``sentences'' (sequences of nodes) via random walks on a graph. 
\item Deep learning methods have also successfully been used to produce node embeddings. Structural Deep Network Embedding (SDNE)~\cite{SDNE} is an example of an autoencoder, a type of artificial neural network that is a commonly used in deep learning to represent complex objects such as images. Graph Convolution Network (GCN)~\cite{kipf2016semi} and GraphSAGE~\cite{GraphSAGE} recursively extract and aggregate some important features similarly to the approach of the Recursive Feature Extraction (ReFeX)~\cite{ReFeX}. See~\cite{zhou2020graph,wu2020comprehensive} for two sample comprehensive surveys (from the many available) on Graph Neural Networks (GNN).
\end{itemize}
Most of the algorithms fall into one or more of the categories defined above but some propose a different approach. One such algorithm is LINE~\cite{LINE}, which explicitly defines two functions to encode the first and the second order proximity.

\subsubsection*{Hyperbolic Spaces}

Some recent research show that, when embedding real-world graphs with scale-free or hierarchical structure, graph distances between pairs of nodes could not be accurately estimated based on the Euclidean distances between the corresponding pairs of embeddings. However, it turns out that embeddings in hyperbolic geometries seem to produce smaller distortion and so they offer a possible alternative for such families of graphs~\cite{boguna2021network}. Currently, they seem to be slower than other methods~\citep{zhang2021systematic} and so not suitable for large-scale networks but it might change in the near future.

In order to see the power of hyperbolic spaces, note that one may approximate tree (discrete) distances arbitrarily well while still being in a continuous space. One cannot hope to approximate this behaviour in Euclidean spaces, but this can be done in the hyperbolic ones, and with only two dimensions~\cite{sala2018representation}. Moreover, it is possible to generalize important algorithms such as multidimensional scaling (MDS) and principal components analysis (PCA) to hyperbolic spaces.

The Hyperbolic Graph Convolutional Neural Network (HGCN)~\cite{HGCN} is the first inductive hyperbolic counterpart of GCN that leverages both the expressiveness of GCNs and hyperbolic geometry to learn inductive node representations for hierarchical and scale-free graphs. Their flexibility (in particular, the curvature of the space) have been successfully leveraged in various areas such as computer vision, NLP, and computational biology. In the context of social networks, they are currently used in collaborative filtering where the goal is to use past user-item/advertisement interactions to build the recommender systems (see, for example,~\cite{HGCF}\footnote{\texttt{https://github.com/layer6ai-labs/HGCF}}).

\subsubsection*{Signed Networks}

The vast majority of existing node embedding algorithms are designed for social networks without sign, typically only with positive links. However, in many social media platforms one may extract both positive and negative links, yielding signed networks. These additional information can be given explicitly; for example, a social news website \texttt{Slashdot.org} allows their users to specify other users as friends or foes, \texttt{Epinions.com} (currently \texttt{Shopping.com}) allows users to mark their trust or distrust to other users on product reviews. Alternatively, one may try to infer whether a given link is positive or negative by investigating the interaction between the two users (comments they produce, likes/dislikes of similar products or posts, etc.). This brings both challenges and opportunities for signed network embedding. While extracting such additional information seems challenging, some initial approaches and experiments indicate some potential, in particular, in link prediction algorithms. A few algorithms have been already proposed: SiNE~\cite{SiNE}, SNE~\cite{yuan2017sne}, SNEA~\cite{wang2017attributed}, ROSE~\cite{javari2020rose}.  Finally, let us mention about a hyperbolic node embedding algorithm for signed networks that was recently explored in~\cite{song2021hyperbolic}.

\subsubsection*{Potential Research Direction \# $\ccc$ (Embedding Sequences of Graphs)}

Node embeddings extract important features of the nodes of the graph. In the context of social networks, they may be representing user's interests, beliefs, emotions, geographic location, age, gender, and many other demographic characteristics. In dynamic scenarios, with users joining/leaving the network, establishing new/deleting old connections, one may want to use embeddings to better understand how communities form/split, how/when/why users become segregated/polarized, who/when/why becomes central and powerful, etc. As a result, temporal graphs become an increasingly important object of study~\citep{hamilton2017representation} and it is expected that extending node embedding techniques to include dynamic aspects will open up various exciting applications.

Suppose that we have snapshots of some network available at various times. Of course, one may embed each of them independently. Unfortunately, most of the embedding algorithms are randomized and so even if the very same graph is embedded twice, the two resulting embeddings will be completely different despite the fact that they try to preserve information from the same graph. Deterministic algorithms also do not guarantee that two similar graphs (say, measured via the edit distance, a measure of similarity, or dissimilarity, between two graphs) yield similar embeddings. In order to be able to use embeddings to better understand/predict dynamics, one needs to couple embeddings of any two consecutive snapshots such that they are similar to each other. For example, one may insist that vectors associated with nodes are not shifted by more than $\delta=\delta(\eps)$ away between the two consecutive snapshots $G_i$, $G_{i+1}$, provided that the edit distance between the two graphs is equal to $\delta |E(G_i)|$. 

Unfortunately, almost all existing node embedding algorithms are inherently transductive. As a result, when a new data point is added to the dataset, then one has to re-run the algorithm from the beginning to train the model. Some of them may be adjusted to inductive setting but they require additional rounds of gradient descent before prediction for new nodes can be used and so such adjustments are computationally expensive. GraphSAGE~\cite{GraphSAGE} is a rare example of graph convolutional networks (GCN) extended to the task of inductive unsupervised learning. For a few more recent attempts to deal with dynamic graphs see Section~4.3 in~\cite{zhou2020graph} and~\citep{huang2020modeling}. 

\subsubsection*{Potential Research Direction \# $\ccc$ (Multi-Layered Graphs)}

In some applications, we are provided with $\ell$ graphs $G_i = (V_i , E_i )$, $1 \le i \le \ell$, with overlapping sets of nodes. We may view it as one graph on the set of nodes $V = \bigcup_{i=1}^{\ell} V_i$ consisting of multiple ``layers''. This is a typical situation in the context of social networks where we often have access to datasets from more than one social platform or there are some auxiliary graphs constructed based on, for example, similarly between users implied by the posts they write and/or read. 

The easiest approach would be to learn features in such multi-layered networks either by treating each layer independently of other layers, or by aggregating the layers into a single (weighted) network on the set of nodes $V$. However, as expected, ignoring the existence of multi-layered structure affects the topological and structural properties of such complex networks. As a result, the importance of individual nodes is altered and cannot be recovered from the simplified picture of the network, leading to wrong identification of versatile nodes and overestimating the importance of marginal nodes~\cite{dedomenico2014,dedomenico2015,dedomenico2016}---see also the book~\cite{bookMultiplex} which ``provides a summary of the research done during one of the largest and most multidisciplinary projects in network science and complex systems'' (Mathematical Review Clippings).

Despite the fact that we know more about processes shaping multi-layered networks, there are still only a few known dedicated embedding algorithms. One of them is OhmNet algorithm~\cite{Zitnik2017} that builds on recent success of unsupervised representation learning methods based on neural architectures. This algorithm uses a form of structured regularization suitable for multi-layer networks. It was tested for the human protein–protein interaction (PPI) network but the ideas and approaches might be potentially useful for multi-layered social networks. 

\subsection{Evaluating Node Embeddings}\label{sec:evaluating_embeddings}

As mentioned in the previous section, there are 100+ node embedding algorithms, most of them have a number of hyper-parameters that one needs to carefully tune for a given task and a given network at hand. Moreover, most algorithms are randomized and not so stable which means that even if the algorithm is run twice on the same network and with the same set of parameters, the resulting embeddings might be substantially different. Some embedding algorithms seem to be performing better than others (for example, in the experiments performed recently in~\citep{embeddings_tests}, Node2Vec worked relatively well for both real world networks as well as synthetically generated ones) but there is no universal choice for all potential applications. As a result, evaluating graph embedding algorithms is a challenging task, typically requiring ad-hoc experiments and tests performed by the domain experts. 

However, in the recent paper~\cite{kaminski2020unsupervised}, a ``divergence score'' was proposed that can be assigned to outcomes of the embedding algorithms to help distinguish good ones from bad ones. This general framework provides a tool for an unsupervised graph embedding comparison and is available at the GitHub repository\footnote{\texttt{https://github.com/ftheberge/Comparing\_Graph\_Embeddings}}. A good embedding should be able to recover (to certain degree, of course) the structure of the network. For example, it is natural to expect that if two vertices are embedded at points that are far away from each other, then the chance that they are adjacent in the graph is smaller compared to another pair of vertices that are close to each other. Embeddings with such property are naturally suitable for link-prediction algorithms. On the other hand, it might be desirable that the communities present in the original network should be clearly separated in the embedded space, and dense sets of points should induce dense graphs spanned on the associated nodes. Embeddings with these properties are suitable for node classification as well as community detection algorithms. 

The authors of~\cite{kaminski2020unsupervised} generalized the classical Chung-Lu model~\cite{ChungLu2006book} to incorporate geometry. Their Geometric Chung-Lu model is then used as the null-model to evaluate the quality of the competing embeddings. Indeed, the model generates a random graph with a given position of nodes (a given embedding under evaluation) and the expected degree distribution. This time the goal is to tune the parameters of the model such that the desired properties are \emph{not} ``surprising'' so that the conclusions derived from the embedding adequately reflect the properties of the network. The initial framework focused on preserving global densities between and within communities but it is currently being adjusted to maximize link predictive power of the embeddings, and to include weighted and/or directed graphs\footnote{\texttt{https://github.com/KrainskiL/CGE.jl}}. 

\begin{figure}[ht]
\begin{center}
\includegraphics[width=7cm]{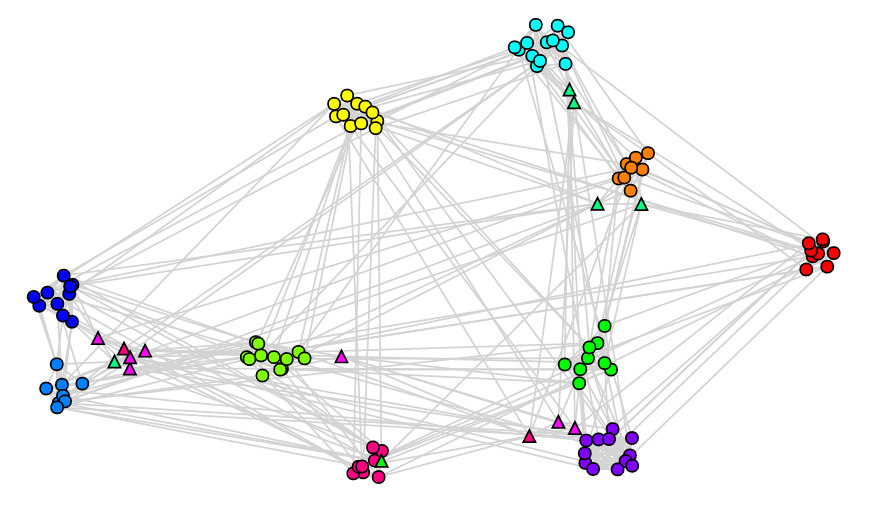}
\includegraphics[width=7cm]{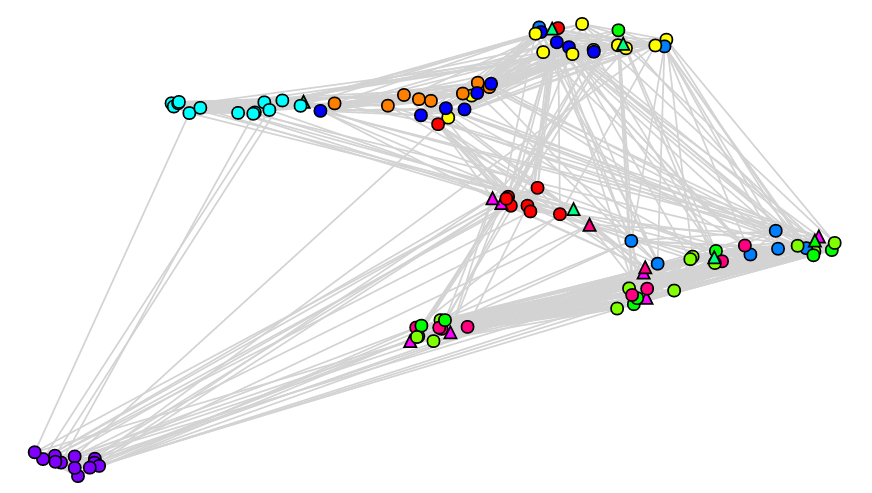}
\end{center}
\caption{The best (left) and the worst (right) scoring embedding of the College Football Graph.}
\label{fig:foot_best}
\end{figure}

In order to illustrate the power of the framework, we present an experiment from~\citep{embeddings_tests} on another well-known real-world network with known community structure, namely, the College Football graph. This graph represents the schedule of United States football games between Division IA colleges during the regular season in Fall 2000~\cite{Girvan7821}. The teams are divided into conferences containing 8--12 teams each. In general, games are more frequent between members of the same conference than between members of different conferences, with teams playing an average of about seven intra-conference games and four inter-conference games in the 2000 season. There are a few exceptions to this rule, as detailed in~\cite{lu2018community}: one of the conferences is really a group of independent teams, one conference is really broken into two groups, and 3 other teams play mainly against teams from other conferences. We refer to those as {\it outlying} nodes, which we represent with a distinctive triangular shape. In Figure~\ref{fig:foot_best}, we show the best and worst scoring embeddings based on the divergence score. The colours of nodes correspond to the conferences, and the triangular shaped nodes correspond to outlying nodes as observed earlier. The communities are very clear in the left plot while in the right plot, only a few communities are clearly grouped together. In order to produce a low dimensional representation of high dimensional data that preserves relevant structure, the Uniform Manifold Approximation and Projection (UMAP\footnote{\texttt{https://github.com/lmcinnes/umap}})~\citep{mcinnes2018umap} was used, a novel manifold learning technique for dimension reduction.

When using the divergence score from the framework as described above, we try to preserve the global structure of the network. This can be extended, as described in~\citep{ExtendedFramework}, by computing an additional score which evaluates embeddings by considering local properties of the network. The best embedding, if needed, can be selected in an unsupervised way using both scores, or the framework can identify a few embeddings that are worth further investigation. This extended framework is flexible, scalable, and can deal with undirected/directed, weighted/unweighted graphs.

\subsubsection*{Potential Research Direction \# $\ccc$ (Selecting an Appropriate Embedding for a Given Task at Hand---Supervised  vs.\ Unsupervised Approach)}

Selecting the best embedding can be done via unsupervised or supervised way. In order for the unsupervised approach to perform well, it is crucial to select an appropriate null-model for a given task. The framework mentioned above allows for tuning the selection process toward community detection algorithms or link prediction algorithms. The geometric Chung-Lu model is used as the null-model since it incorporates two important aspects, geometry as well as the degree distribution, but is simple enough so that one can easily compute the probability of two nodes to be adjacent in the model. Moreover, all edges are generated independently which allows one to investigate the desired properties and use them to benchmark embeddings. However, it is clearly not the only reasonable model to use. For a specific task at hand, or if some additional information about the graph or the embedding is provided, one may choose to use some other, more appropriate, model. In particular, it might be good to experiment with various random hyperbolic geometric graphs, especially if hierarchical networks are under investigation. Despite the fact that typically null-models are easy, let us mention that it is not a problem if the model is too complex for the desired properties to be theoretically analyzed---indeed, one may always use simulations to investigate them. Finally, when labelled training data becomes available, one may alternatively use (or complement the unsupervised ``divergence score'') supervised learning tools to benchmark the embeddings available and select the one that scores the best for a given task at hand. 

\subsection{Community Detection}

A network has community structure if its set of nodes can be split into subsets that are densely internally connected. For example, in social networks, communities may represent user interests, occupation, age, etc. This is a very important (unsupervised) task when analyzing complex networks. Indeed, after finding communities one can better understand the role of each node, their interactions, and it allows one to focus on relevant portions of the graph, to name only a few possible applications. The first problem we consider is node partitioning, where we seek to divide the set of $n$ nodes into $k$ non-overlapping subsets, where each subset yields a dense community subgraph. Note that the number of communities (that is, the value of $k$) is generally unknown, and the number of possible partitions is enormous even for small graphs. 

There are several natural ways to define ``dense subgraph'' to make the task of finding communities well-defined but the general idea is that the number of internal connections (edges) should be larger than the expected number of connections based on some global statistics of the whole graph.
One commonly used measure that is used to find communities is the modularity function~\cite{modularity}, which is based on the comparison between the actual density of edges inside a community and the density one would expect to have if the nodes of the graph were attached at random, without any community structure, while respecting the nodes’ degrees. 

Several graph clustering algorithm aim at maximizing the modularity function, one of the best ones being the well-known Louvain algorithm~\cite{louvain}.
In this algorithm, small communities are first found by optimizing modularity locally on all nodes. This yields the level-1 partition of the nodes.
Then, each small community is grouped into one node and the original step is repeated on this smaller graph, yielding the next level of partition. The process is repeated until no improvement on the modularity function can be achieved. As a result, a hierarchy of partitions is obtained with decreasing granularity.

The Louvain algorithm offers good trade-off between the quality of the clusters it produces and its speed but it has some stability and resolution issues.
Instability is due to the randomization of the node ordering when locally optimizing the modularity, which can lead to very different partitions when running the algorithm multiple times on the same graph. 
Resolution issue is an issue with modularity itself~\cite{resolution} and it can lead to the merger of small communities. 
Such behaviour can be addressed by trying to ``break up'' the communities individually or by increasing aversion of nodes to form communities by adding self-loops or modifying the modularity function directly.

\begin{figure}[htb]
\begin{center}
\includegraphics[width=9cm]{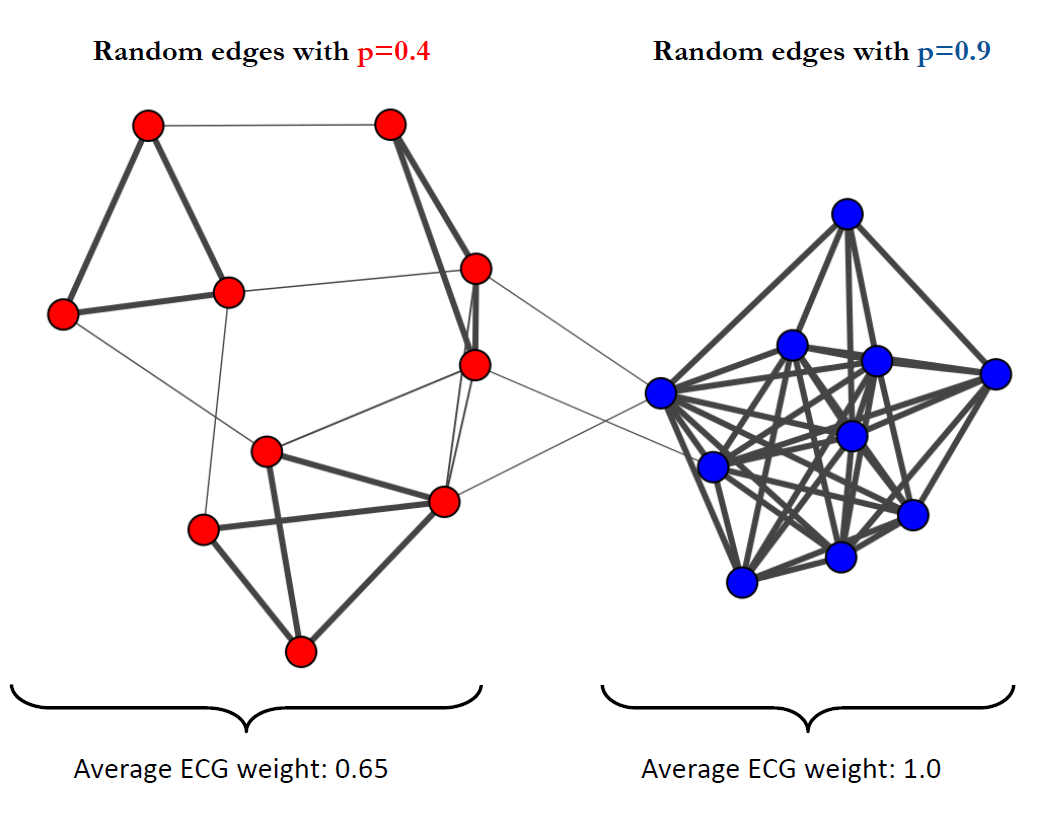}
\end{center}
\caption{Clustering a small graph with two communities: weak (red, edge density 0.4) and strong (blue, edge density 0.9) using ECG. We see that ECG derived weights (thick lines have high weights) are useful in identifying tight communities and sub-structures. }
\label{fig:ecg}
\end{figure}

Another way to address the above issues is by using an ensemble of partitions, as proposed with the Ensemble Clustering for Graphs (ECG) algorithm~\cite{ecg}. 
With ECG, we start by running the level-1 of the Louvain algorithm several times, thus building an ensemble of highly granular partitions. For each edge, the number of times it is internal (that is, both nodes are within the same community) for the partitions in the ensemble is used as edge weight to obtain the final partition using the Louvain algorithm.
Thus, not only are the obtained partitions more stable and often of better quality than with the ``vanilla'' Louvain, but the ECG derived edge weights are useful to assess the quality of the clusters we get.
We illustrate this is Figure~\ref{fig:ecg}, where we ran ECG on a small network with two communities: red nodes form a weakly connected community (40\% of pairs have an edge) and blue nodes form a tight community (90\% of pairs have an edge). We also report the edges with high ECG weights with thicker black lines. 
All edges in the blue community have maximal ECG weight of 1, which is not the case for the red community, but in this case, we do see that the tight sub-structures (triangles) do have higher ECG weights. 

There are several other graph clustering algorithms (see for example~\cite{fortunato2010}, or~\cite{yang2016}). For example, the Leiden algorithm~\cite{leiden} is a proposed improvement to the Louvain algorithm fixing issues such as disconnected communities; note that Leiden can be used within the ECG code\footnote{github.com/ftheberge/graph-partition-and-measures}.
CNM~\cite{cnm} is a hierarchical, agglomerative algorithm based on the modularity function. Other hierarchical algorithms include Ravasz agglomerative algorithm~\cite{ravasz} based on the topological overlap, and the Girvan-Newman divisive algorithm~\cite{girvan} which is based on edge betweenness centrality.
Other algorithms include label propagation~\cite{label}, spectral bisection~\cite{spectral} and Infomap~\cite{infomap}, to name a few.
For more details, we refer the reader to, for example,~\citep{Kaminski2021book}.

\subsubsection*{Potential Research Direction \# $\ccc$ (More General Community Detection and Using Several Sources of Information)}

Graph partitioning to find communities is a very useful tool in network science but, depending on the problem at hand, one or several of the following generalizations may be of relevance.
\begin{itemize}
    \item Some nodes could be allowed to be excluded from communities (so-called ``noise'').
    \item Nodes could be part of several communities; this is the overlapping communities problem.
    \item The same nodes could interact over several distinct networks, for example, Twitter, Facebook, and Reddit.
\end{itemize}

Ideas based on ensemble clustering as in ECG could be useful to address the generalizations above. For example, ECG derived weights could be used to identify ``noise'' nodes after removing weak edges. Similarly, nodes having strong edges spanning several communities could indicate overlapping clusters with these nodes belonging to more than one cluster. There are some known methods for finding overlapping communities such as clique percolation~\cite{cpm}, which looks for overlapping cliques (small fully connected subgraphs such as triangles), ego-splitting~\cite{ego}, where nodes can be duplicated to properly reflect their multiple community membership, and edge clustering~\cite{edge_cluster}.

Independently, one may try to use some external information to improve the quality or the stability of the clustering algorithm. 
\begin{itemize}
    \item Extra information could be available for the nodes; for example if those represent people interacting over social networks, we may know about the country of origin, language(s) used, interests, text generated, etc.
    \item Extra information could be available for the edges; for example the topic of a message sent, the sentiment (such as like or dislike), the timestamp(s), etc.
\end{itemize}

Such metadata could, for example, be used to obtain an embedding of the nodes, followed by some clustering of the points to obtain several tight clusters. This, in turn, can be used as starting point for a Louvain-like algorithm to improve its quality. If several sources of metadata are available, then an ensemble methods could be tried. Finally, another possibility is to modify the modularity function itself to favour clustering of nodes such that the corresponding points exhibit high level of similarity derived from the external information available for the nodes and edges.

\subsection{Hypergraphs}\label{sec:hypergraphs}

Real-world complex networks are usually being modelled as graphs. The concept of graphs assumes that the relations between nodes within the network are binary; however, this is not always true for many real-life scenarios, including social network interactions. For example, a group of users commenting on the same post or liking/disliking a given picture typically consists of more than two users.

Hypergraph $H=(V,E)$ consists of the set of nodes $V$ and the set of hyperedges $E$; each hyperedge $e \in E$ is a subset of $V$, but not necessarily of size 2 as in the case of graphs. As a result, it is a natural generalization of graphs. More importantly, many complex networks that are currently modelled as graphs would be more accurately modelled as hypergraphs. This includes the collaboration network in which nodes correspond to researchers and hyperedges correspond to papers that consist of nodes associated with researchers that co-authorship a given paper. Similarly, social networks can be modelled as hypergraphs with hyperedges consisting of all users that interact with each other by commenting on the same post/article. However, here the situation is even more complex. Indeed, this type of data can be more accurately modelled by a dynamic tree structure with the root associated with a given post/article followed by a tree of comments/likes/dislikes.

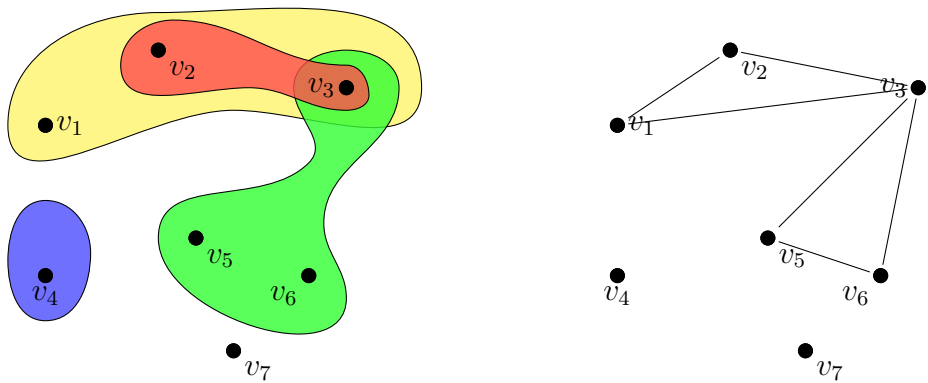
\begin{figure}[ht]
\begin{center}
\begin{tikzpicture}
    \node (v1) at (0,2) {};
    \node (v2) at (1.5,3) {};
    \node (v3) at (4,2.5) {};
    \node (v4) at (0,0) {};
    \node (v5) at (2,0.5) {};
    \node (v6) at (3.5,0) {};
    \node (v7) at (2.5,-1) {};

    \begin{scope}[fill opacity=0.8]
    \filldraw[fill=yellow!70] ($(v1)+(-0.5,0)$) 
        to[out=90,in=180] ($(v2) + (0,0.5)$) 
        to[out=0,in=90] ($(v3) + (1,0)$)
        to[out=270,in=0] ($(v2) + (1,-0.8)$)
        to[out=180,in=270] ($(v1)+(-0.5,0)$);
    \filldraw[fill=blue!70] ($(v4)+(-0.5,0.2)$)
        to[out=90,in=180] ($(v4)+(0,1)$)
        to[out=0,in=90] ($(v4)+(0.6,0.3)$)
        to[out=270,in=0] ($(v4)+(0,-0.6)$)
        to[out=180,in=270] ($(v4)+(-0.5,0.2)$);
    \filldraw[fill=green!80] ($(v5)+(-0.5,0)$)
        to[out=90,in=225] ($(v3)+(-0.5,-1)$)
        to[out=45,in=270] ($(v3)+(-0.7,0)$)
        to[out=90,in=180] ($(v3)+(0,0.5)$)
        to[out=0,in=90] ($(v3)+(0.7,0)$)
        to[out=270,in=90] ($(v3)+(-0.3,-1.8)$)
        to[out=270,in=90] ($(v6)+(0.5,-0.3)$)
        to[out=270,in=270] ($(v5)+(-0.5,0)$);
    \filldraw[fill=red!70] ($(v2)+(-0.5,-0.2)$) 
        to[out=90,in=180] ($(v2) + (0.2,0.4)$) 
        to[out=0,in=180] ($(v3) + (0,0.3)$)
        to[out=0,in=90] ($(v3) + (0.3,-0.1)$)
        to[out=270,in=0] ($(v3) + (0,-0.3)$)
        to[out=180,in=0] ($(v3) + (-1.3,0)$)
        to[out=180,in=270] ($(v2)+(-0.5,-0.2)$);
    \end{scope}

    \foreach \v in {1,2,...,7} {
        \fill (v\v) circle (0.1);
    }

    \fill (v1) circle (0.1) node [right] {$v_1$};
    \fill (v2) circle (0.1) node [below right] {$v_2$};
    \fill (v3) circle (0.1) node [left] {$v_3$};
    \fill (v4) circle (0.1) node [below] {$v_4$};
    \fill (v5) circle (0.1) node [below right] {$v_5$};
    \fill (v6) circle (0.1) node [below left] {$v_6$};
    \fill (v7) circle (0.1) node [below right] {$v_7$};

\end{tikzpicture}
\hspace*{2cm}
\begin{tikzpicture}
    \node (v1) at (0,2) {};
    \node (v2) at (1.5,3) {};
    \node (v3) at (4,2.5) {};
    \node (v4) at (0,0) {};
    \node (v5) at (2,0.5) {};
    \node (v6) at (3.5,0) {};
    \node (v7) at (2.5,-1) {};

    \draw (v1) to (v2);
    \draw (v2) to (v3);
    \draw (v1) to (v3);

    \draw (v5) to (v6);
    \draw (v5) to (v3);
    \draw (v3) to (v6);

    \foreach \v in {1,2,...,7} {
        \fill (v\v) circle (0.1);
    }

    \fill (v1) circle (0.1) node [right] {$v_1$};
    \fill (v2) circle (0.1) node [below right] {$v_2$};
    \fill (v3) circle (0.1) node [left] {$v_3$};
    \fill (v4) circle (0.1) node [below] {$v_4$};
    \fill (v5) circle (0.1) node [below right] {$v_5$};
    \fill (v6) circle (0.1) node [below left] {$v_6$};
    \fill (v7) circle (0.1) node [below right] {$v_7$};

\end{tikzpicture}
\end{center}
\caption{Hypergraph (left) and its 2-section (right).}
\label{fig:hypergraph_2_section}
\end{figure}

Unfortunately, the theory and tools are still not sufficiently developed to allow most problems (including community detection algorithms and centrality measures) to be tackled directly within this context. As a result, researchers and practitioners often create the 2-section graph of a hypergraph of interest, that is, replace each hyperedge with a clique---see Figure~\ref{fig:hypergraph_2_section}. After moving to the 2-section graph, one clearly loses some information about hyperedges of size greater than two and so there is a common believe that one can do better by using the knowledge of the original hypergraph. Having said that, there has been a recent surge of interest in higher-order methods, especially in the context of hypergraph clustering which we will concentrate on from now on. See~\cite{battiston2020networks} for a recent survey on the higher-order architecture of real complex systems, and~\cite{feng2019hypergraph} for a hypergraph neural networks framework (HGNN) for data representation learning.

Graph clustering is the task of partitioning the set of nodes of a given graph into clusters such that clusters are substantially denser than the global density of a network (see the two surveys, by Fortunato and Hric~\cite{Fortunato_survey}, and by Schaeffer~\cite{Schaeffer_survey}). Since there are various ways to partition the nodes of a hyperedge, there are numerous generalizations of a graph-based objective function to hypergraphs. Various notions of hyperedge cuts have been considered in the past (see references in~\cite{Gleich}). Recently, a hypergraph clustering objective was proposed in~\cite{Gleich} that differently treats hyperedges and pairwise edges in a parametric fashion.

\subsubsection*{Potential Research Direction \# $\ccc$ (Hypergraph Modularity Function)}

There are some recent attempts to deal with hypergraphs in the context of graph clustering. Kumar \emph{et al.}~\cite{kumar2019new,kumar2020hypergraph} still reduce the problem to graphs but use the original hypergraphs to iteratively adjust weights to encourage some hyperedges to be included in some cluster but discourage other ones. Moreover, in~\cite{kaminski2019clustering} a number of extensions of the classic null model for graphs are proposed that can potentially be used by true hypergraph algorithms. 

Unfortunately, there are many ways such extensions can be done depending on how often nodes in one community share hyperedges with nodes from other communities. Fortunately, it is possible to unify all variants of the graph modularity function to hypergraphs and put them into one framework. Two prototype algorithms investigated in~\cite{kaminski2020community} show the potential of the framework but its true power needs to be confirmed by experiments on larger networks. The authors of that paper still work on designing a scalable algorithm. 

Finally, let us stress it again that hypergraphs are just the beginning and there are many ways to generalize hypergraphs even further. One important generalization is the one in which nodes are assigned roles within each hyperedge. Many complex networks provide such additional information: research articles have junior and senior authors, political bills have sponsors and supporters, movies have starring and supporting actors, e-mails have senders, receivers, and carbon copies, etc. In order to model these networks, annotated hypergraphs were introduced in~\cite{chodrow2020annotated} as natural polyadic generalizations of directed graphs. To facilitate data analysis with annotated hypergraphs, the authors of that paper construct a role-aware configuration null-model for these structures and prove an efficient Markov Chain Monte Carlo scheme for sampling from it. In order to take advantage of this additional information, one should adjust the modularity function once again to incorporate this new null-model. 

\subsection{Understanding the Dynamics of Networks} 

One of the very first model of dynamic network is the preferential attachment model introduced in the paper of Barab\'asi and Albert~\cite{barabasi1999emergence} who observed a power law degree sequence for a subgraph of the World Wide Web; soon after the same property was observed for the internet graph~\cite{faloutsos2011power}. It is an important model as it explains why power-law degree distribution occurs in many real-world complex networks. Indeed, this model was rigorously analyzed in~\cite{bollobas2011degree,bollobas2004diameter} and the conclusion is as follows. The existence of a power-law degree distribution is a consequence of a rich-get-richer phenomenon: the preferential attachment rule incorporated in the model makes new nodes to be more likely to connect to the more connected nodes than to the smaller nodes, creating as a result power-law degree distribution. 

One drawback of the above mentioned model is that there is a strong correlation between the age of a node and its degree, making it impossible for young nodes to gain a substantial number of neighbours. The fitness model introduced in~\cite{bianconi2011competition} assigns a fitness to new born nodes that corresponds to some intrinsic property that propels them ahead of the pack.  Another problem of the original model is that it generates power-law degree distribution but the exponent cannot be easily tuned. This issue was addressed in~\cite{ostroumova2013generalized} and in the very recent work, a preferential rule was adjusted to mimic any empirical degree distribution~\cite{giroire2020random}. 

A model is said to have accelerating growth if the number of edges grows non-linearly with the number of vertices. Such models became important as there is an evidence of increasing edge density (densification) and decreasing diameter in existing networks~\cite{leskovec2005graphs,leskovec2007graph}. Among networks found by the authors to exhibit increasing average out-degree over time were: ArXiv citation, patent citation and autonomous systems (internet routers). The authors of that paper were concerned to find causal models of densification, and propose explanatory mechanisms for this, such as community guided attachment, and the forest fire model. In particular, simulations of the forest fire model show that densification itself does not necessarily cause the diameter to shrink. Another accelerating growth model is the preferential attachment model in which the degrees of newly added nodes increase over time~\cite{cooper2011scale}.

There are many other models that explain the occurrence of various typical properties of complex networks. The Watts--Strogatz model produces graphs with short average path lengths and large clustering coefficient~\cite{watts1998collective}. There are also many geometric models of interest, one of them is the Spatial Preferential Attachment model that combines the rich-get-richer phenomenon with geometrical aspects~\cite{aiello2008spatial,cooper2014some}.
For more examples of such models see, for example,~\cite{dorogovtsev2013evolution}.

Another type of models have the purpose to represent network dynamics on the basis of observed longitudinal data, and evaluate these according to the paradigm of statistical inference. Simulation Investigation for Empirical Network Analysis (SIENA)\footnote{\texttt{https://www.stats.ox.ac.uk/\~{}snijders/siena/}} is a software to preform the statistical estimation of models for repeated measures of social networks according to the Stochastic Actor-oriented Model. For more details, see a review article~\cite{snijders2017stochastic} or the manual for SIENA 4.0\footnote{\texttt{https://www.stats.ox.ac.uk/\~{}snijders/siena/RSiena\_Manual.pdf}} compiled recently (May 11, 2021). This model has an ``actor-oriented'' nature which means that it models change from the perspective of the actors (nodes). According to the manual, the model is suitable for the analysis of:
\begin{itemize}
\item the evolution of a directed or non-directed one-mode network (e.g., friendships in a classroom),
\item the evolution of a two-mode network (e.g., club memberships in a classroom: the first mode is constituted by the students, the second mode by the clubs),
\item the evolution of an individual behaviour (e.g., smoking), and
\item the co-evolution of one-mode networks, two-mode networks and individual behaviours (e.g., the joint evolution friendship and smoking; or of friendship and club membership).
\end{itemize}
An important drawback is that this method is applicable to only very small graphs consisting of approximately 10 to 1,000 nodes. However, some ideas and approaches might be useful for large graphs.

Another general modeling framework for temporal network data analysis is to extend Exponential Random Graph Models (ERGMs)~\cite{harris2013introduction} to Temporal Exponential family Random Graph Models (TERGMs)~\cite{leifeld2018temporal}\footnote{\texttt{http://statnet.org/Workshops/tergm\_tutorial.html}}. For example, Separable Temporal ERGMs (STERGMs) are an extension of ERGMs for modeling dynamic networks in discrete time, introduced in~\cite{krivitsky2014separable}\footnote{\texttt{https://cran.r-project.org/web/packages/tergm/vignettes/STERGM.pdf}}. Within this framework, one may obtain maximum-likehood estimates for the parameters of a specified model for a given data set; simulate additional networks with the underlying probability distribution implied by that model; test individual models for goodness-of-fit, and perform various types of model comparison.

\medskip

Let us now briefly discuss a few aspects that seems to be important in the context of dynamics of social networks. 

\subsubsection*{Human-bot Interaction and Spread of Misinformation}

Social networks created an information platform in which automated accounts (both human as well as bots---software-assisted accounts) can try to take advantage of the system for various opportunistic reasons: trigger collective attention~\cite{lehmann2012dynamical,de2020unraveling}, gain status~\cite{cha2010measuring,stella2019influence}, monetize public attention~\cite{carter2016hustle}, diffuse disinformation~\cite{bail2020assessing,freelon2020black}, or seed discord~\cite{woolley2018computational}. 

It is known that a high fraction of active Twitter accounts are bots and that they are responsible for much disinformation. We better understand the role they play in the diffusion of false information. In particular, it seems that they play an important role at the initial stage of diffusion by amplifying low-credibility content but they cannot distinguish between true and false, that is, real human accounts are more likely to spread false news. (See~\cite{gonzalez2021bots} and the references there.) Indeed, it is possible to characterize the peculiar behaviour of certain individuals who massively use bots to enhance their online visibility and influence. The term \emph{cyborg} or \emph{augmented human} has been used in this context to identify, indistinctly, bot-assisted human or human-assisted bot accounts. (See~\cite{stella2019influence} and the references there.) It is also worth mentioning that spreading information and banning strategies depend on whether social platform is moderated or not~\cite{artime2020effectiveness}. 
In a recent paper, a framework for learning/opinion formation of an individual from signed network data is proposed~\citep{meng2021whom}. This framework can be used to understand, for example, why people end up trusting misinformation sources. 
Finally, let us mention that various countries have different levels of infodemic risk~\cite{gallotti2020assessing}, adding another level of complexity for a person trying to model these processes.

Despite the fact that we understand social networks better, it is clear that the exact mechanisms responsible for spreading false information (for example, during political events) are still far from being well understood. In fact, there are some recent results that suggest that spreading mechanisms might have indistinguishable population-level dynamics~\cite{hebert2020macroscopic}.

\subsubsection*{Social Bursts in Collective Attention}

We are all flooded with a large amount of information, impossible to consume. Indeed, human attention is a limited resource and a way a given individual reacts to a given information is a complex interplay between individual interests and social interaction. It is known that a collective attention is typically characterized by a quickly growing accumulated focus on a specific topic (for example, presidential elections) until a well identified peak of collective attention is reached. This first phase is followed by the second phase in which one  observes a slow decay of interest. (See~\cite{lehmann2012dynamical} or~\cite{de2020unraveling} and the references there.)

Some initial research neglected the effects of the underlying social structure~\cite{wu2007novelty} but it is clear that underlying network structure plays an important role in the process~\cite{gleeson2016effects}. In particular, the authors of~\cite{de2020unraveling} combine two simple mechanisms to explain the dynamic of the collective attention: a preferential attachment process shaping the network topology and a preferential attention process responsible for individual's attention bias towards specific users of the network.

\subsubsection*{Social Learning (Segregation, Polarization)}

Social learning is a term that refers broadly to the processes by which a person's social environment shapes their actions (how a person behaves) and cognitions (how a person thinks). For example, in a recent paper~\citep{semenova2021reddit} the authors show that asset discussions on WallStreetBets (WSB) are self-perpetuating: an initial set of investors attracts a larger and larger group of excited followers. Based on that, a model for how social contagion impacts prices is developed. Similarly, in the context of social networks a user may adopt the cognitions or behaviours from those they have an opportunity to interact with directly.  At the same time, products of learning also shape the social environments, since individuals also exercise control over their social environment and potentially select network partners as a function of individual attributes, including their behaviours and cognitions~\cite{aral2009distinguishing}. As a result, social learning is a complex process; here we only concentrate on segregation and polarization. 

There are various models that try to explain why and how segregation occurs, starting from the classic model of residential segregation of Schelling~\cite{schelling1969models}. Schelling's results also apply to the structure of networks; namely, segregated networks always emerge even if the users are assumed to have only a small aversion from being connected to others who are dissimilar to themselves, and yet no actor strictly prefers a segregated network~\cite{henry2011emergence}. The power of aversion is often amplified by homophily, a tendency for people to have ties with people who are similar to themselves~\cite{byrne1997overview}. For example,~\cite{bener2016empirical} develops and tests empirical models of how social networks evolve over time; particularly, how people in a social network choose links on the basis of their own attributes, and how individual attributes are in turn shaped by network structure. In~\citep{li2021does}, the authors try to investigate how homophily shapes the topology of adaptive networks (see also a long list of other related models referenced in that paper).

An extreme situation occurs when a group is divided into two opposing sub-groups having conflicting and contrasting positions, goals and points of view, with only a few individuals remaining neutral. A domain where polarization typically occurs is politics but there are other domains that often experience it, such as global warming, gun control, same-sex marriage or abortion. The modularity function mentioned earlier identifies communities but it is known that it is not a direct measure of antagonism between groups~\cite{guerra2013measure}. As a result, other metrics being able to identify and to analyze the boundary of a pair of polarized communities were proposed, which better captures the notions of antagonism and polarization. The presence of polarization changes the dynamics of a network. Indeed, for example it is known that Twitter users are unlikely to be exposed to cross-ideological content from the clusters of users they followed, as these were usually politically homogeneous~\cite{himelboim2013birds}. (See also a recent survey of Twitter research~\cite{antonakaki2021survey}.)

\subsubsection*{Potential Research Direction \# $\ccc$ (Tools Based on the Null-models)}

Apart from the theoretical interest, models of complex networks can have significant impact on the design of practical tools. Indeed, understanding the principles driving the organization and behaviour of such networks proved to be helpful to design various important tools:
sampling algorithms~\cite{leskovec2006sampling},
recommendation systems~\cite{seo2017personalized},
defence systems against attacks from bots~\cite{ferrara2016rise} and spam campaigns~\cite{benevenuto2010detecting},
and measuring of users' influence~\cite{morales2014efficiency}.
Having said that, there is clearly a room for more research in this area. Below we mention a few potential directions.

\textbf{Community Detection}: The Chung-Lu random graph model is currently used to define the modularity function which guides the Louvain algorithm to find communities. It tries to find a partition that induces ``surprisingly'' dense communities in comparison to the underlying null-model. It could also be used to identify overlapping clusters. 

\textbf{Anomalies Detection}: Going down to the level of nodes, the same Chung-Lu random null-model should also be able to identify leaders, followers, and other ``unusual'' members of the community. Initial experiments on the College Football network presented in~\citep{Kaminski2021book} show that this approach has a potential in identifying anomalies. 
There are several other methods for graph-based anomaly detection that could be possibly investigated; see for example \cite{akoglu2015} and \cite{fraud2020}.

\textbf{Link Prediction}: Combining the null-model (in this case, random geometric graph) with a carefully selected embedding of a graph can be a useful tool in link prediction algorithms. Such algorithms should take into account not only positions of the nodes in the embedded space but complement it with some additional information such as community membership and triadic closure (a measure of the tendency of edges to form triangles), to make a better, density based, link prediction. 

\medskip

Many social networks are heterogeneous by nature, that is, there are various types of relationships that are present in these networks. One may combine all of these relationships into a single weighted network but some important information is lost during this process. A few algorithms try to take advantage of heterogeneous information---see, for example,~\cite{sun2011co,sun2012will} for link prediction algorithms---but more work in this area is clearly needed.  

Even less tools try to take advantage of dynamics aspects. On the other hand, for example, understanding how communities are formed should allow us not only to identify current communities but also to predict the future shape of a network. Moreover, simple principles such as rich-get-richer or fitness might be enough to build a model which can predict which nodes are going to be central and influential in the future, and which nodes are losing their power. 

Finally, a good tool should try to combine all possible aspects (dynamics, heterogeneity, higher-order structures, community distribution, metadata, etc.) to ``squeeze the last drop'' from the dataset. There are very few techniques that combine at least 2 of such aspects. One such example is the recent paper~\cite{fard2019relationship} that combines heterogeneity and temporal evolution to make better link predictions.

\subsection{Generating Synthetic Networks}

Many machine learning algorithms and tools are unsupervised in nature, including community detection and anomalies detection algorithms. Unfortunately, these algorithms are often quite sensitive and so they cannot be fine-tuned for a given family of networks we want these algorithms to work on. For example, some algorithms perform well on networks with strong communities but perform poorly on graphs with weak communities; often density or degree distribution affect both the speed as well as the quality of a given algorithm, etc. Because of that it is important to be able to test these algorithms for various scenarios that can only be done using synthetic graphs that have built-in community structure, power-law degree distribution, and other typical properties observed in complex networks. 

\begin{figure}[ht]
\begin{center}
\includegraphics[width=7cm]{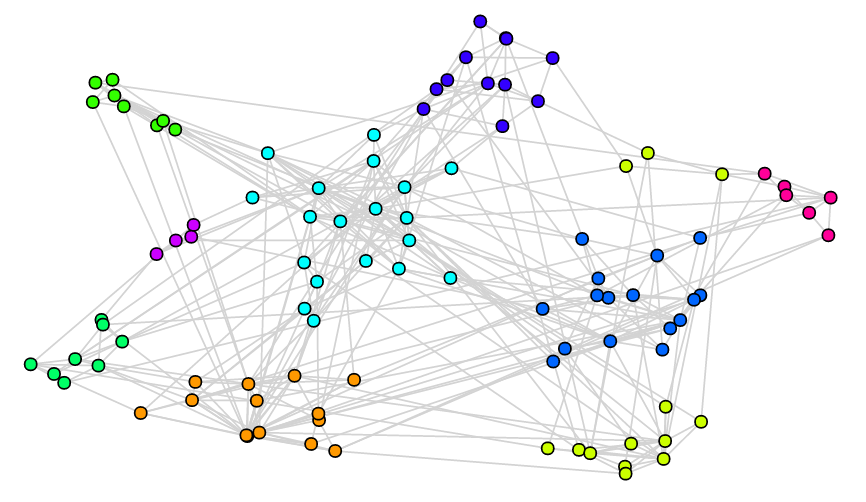}
\includegraphics[width=7cm]{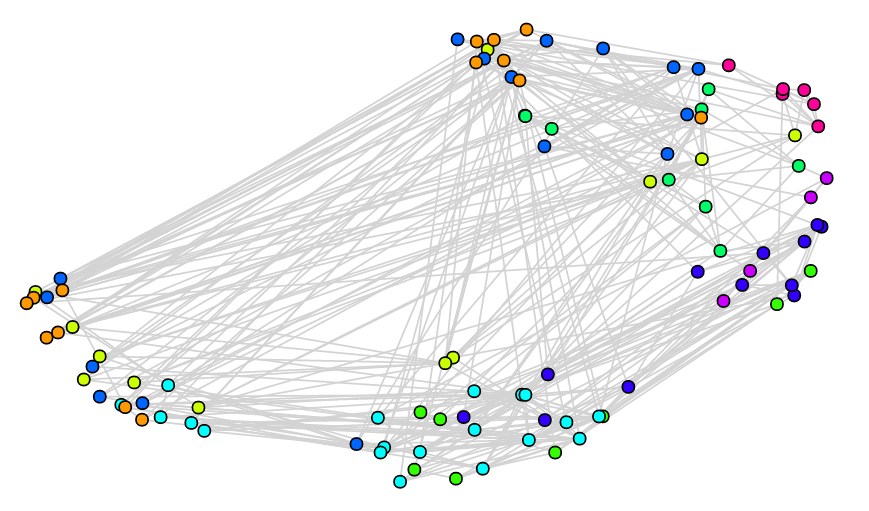}
\end{center}
\caption{The best (left) and the worst (right) scoring embedding of the LFR graph.}
\label{fig:LFR}
\end{figure}

Classical random graphs, such as binomial random graphs or random regular graphs, are very interesting from theoretical point of view~\cite{bollobas2001random,janson2011random,frieze2016introduction} but are not suitable for practical applications. Hence, in order to model the community structure, the Stochastic Block Model (SBM)~\cite{holland1983stochastic} was introduced (see~\cite{funke2019stochastic} for an overview of various generalizations). On the other hand, Chung-Lu model~\cite{ChungLu2006book} was introduced to generate graphs with non-homogeneous degree distributions, including power-law degree distribution that is commonly present in complex networks. The LFR (Lancichinetti, Fortunato, Radicchi) model~\cite{lancichinetti2008benchmark,lancichinetti2009benchmarks} generates networks with communities and at the same time it allows for the heterogeneity in the distributions of both vertex degrees and of community sizes---see Figure~\ref{fig:LFR}. As a result, it became a standard and extensively used method for generating artificial networks. An alternative, ``LFR-like'' random graph model, the Artificial Benchmark for Community Detection (ABCD graph)~\cite{kaminski2020artificial} was recently introduced and implemented\footnote{\texttt{https://github.com/bkamins/ABCDGraphGenerator.jl/}}. The authors of~\cite{kaminski2020artificial} continue working on the model, currently working on the implementation that uses multiple threads (ABCDe)\footnote{\texttt{https://github.com/tolcz/ABCDeGraphGenerator.jl/tree/CFGparallel}}. LFR and ABCD produce graphs with comparable properties but ABCD is faster than LFR (ABCDe is even faster than ABCD) and can be easily tuned to allow the user to make a smooth transition between the two extremes: pure (independent) communities and random graph with no community structure.

\subsubsection*{Potential Research Direction \# $\ccc$ (Generating Synthetic Higher-order Structures)}

As mentioned a few times earlier, many complex networks (including social networks) are better modelled with higher-order structures such as hypergraphs. Synthetic graph models are available but there is a need for more scalable hypergraph models that mimimc the properties of real world networks. (Let us again mention the annotated hypergraphs that were recently introduced in~\cite{chodrow2020annotated} but such models are still rare.) Good models of graphs/hypergraphs with communities (including overlapping communities) and/or anomalous nodes are also in need for a purpose of testing and tuning potential community/anomaly detection algorithms. Finally, it would be good to have a synthetic model that generates a sequence of graphs/hypergraphs to be able to train and benchmark the algorithms that try to capture the dynamic of networks.

\newpage

\section{Conclusions} 

We have presented above a comprehensive survey of generative methods for social media analysis. Its topic is timely and needed, given the recent interest in social media and their role in communications, community building, fake news etc. The survey covers four fundamental areas of social media analytics: 
relevant ontologies and data management aspects, natural language text generation and social media, sentiment analysis in the social media context, and the network approaches to social media understanding. Each of the chapters is rounded up by a discussion of the current limitations and potential future work in the area of the chapter.